\documentclass[12pt]{scrartcl}

\usepackage{graphicx}
\usepackage{multirow} 
\usepackage[version=4]{mhchem} 
\usepackage{fourier} 
\usepackage{FiraSans} 
\usepackage{authblk} 
\usepackage{microtype} 
\usepackage{xcolor} 

\addtokomafont{title}{\large}      
\addtokomafont{author}{\large} 

\title{How Does Intercalation Reshape Layered Structures? A First-Principles Study of Sodium Insertion in Layered Potassium Birnessite}

\author[1,$\dagger$]{Adriana Lee Punaro}
\author[2,$\dagger$]{Daniel Maldonado-Lopez}
\author[1]{Jorge L. Cholula-Díaz}
\author[1,$*$]{Marcelo Videa}
\author[2,3,$*$]{Jose L. Mendoza-Cortes}

\affil[1]{School of Engineering and Sciences, Tecnológico de Monterrey, Monterrey, Nuevo León, Mexico}
\affil[2]{Department of Chemical Engineering and Materials Science, Michigan State University, East Lansing, MI, USA}
\affil[3]{Department of Physics and Astronomy, Michigan State University, East Lansing, MI, USA}
\affil[$\dagger$]{\textit{These authors contributed equally to this study}}
\affil[$*$]{Corresponding authors: mvidea@itesm.mx, jmendoza@msu.edu}

\date{}

\begin{document}

\maketitle

\begin{abstract}
This study presents a first-principles study at the level of hybrid-level density functional theory of the sodium intercalation process in a layered potassium birnessite (a layered manganese dioxide, $\delta$-\ce{MnO2}). Understanding the intercalation processes of $\delta$-\ce{MnO2} is a vital step in advancing its potential innovative applications. Through a formation energy formalism, we analyze the stability of the structure as sodium ions (Na$^+$) are intercalated between layers. Simulated Raman spectra allow us to find relationships between the vibrational and structural properties of the material, i.e. we identify the most important vibrational modes and related them to the structural/geometrical change. The diffusion of \ce{Na+} and \ce{K+} ions in birnessite is studied by transition state theory, determining the energy barriers to ion displacement in the interlayer. The symmetry and planar density of the system are characterized by simulated X-ray diffraction and geometrical analysis of the optimized structures. Through binding energy analysis, we also find that the \ce{Na+} ions are more loosely bound to the lattice as they reach the saturation limit. Finally, the electronic properties are studied via spin-polarized densities of states. As intercalants are added, the electronic properties are profoundly modified, resulting from modification of Mn oxidation states, lattice distortions, and symmetry effects. Moreover, some of the intercalated structures behave as bipolar magnetic semiconductors with potential applications in spintronics devices. In other words, the band gaps and magnetic behavior of the system can be controlled by intercalation. This work provides an overarching analysis of intercalated birnessite and describes the essential properties of potassium birnessite and co-intercalation with Sodium as a next-generation energy, electronic, and spintronic material.

\end{abstract}

\section{Introduction}
In a technology-driven world, there is a constant need for the development of low-cost, high-energy storage, and semiconducting materials. Among the materials studied in energy research, the layered birnessite manganese dioxide ($\delta$-\ce{MnO2}) has emerged as a promising candidate for these technologies. Birnessite’s two-dimensional structure, ability to store ions in the interlayer, high capacitance, and Mn$^{3+}$/Mn$^{4+}$ reversible redox activity, make it an attractive material for next-generation energy storage and conversion systems. Its pseudocapacitive charge storage properties, high power and energy density qualify it as a potential material for supercapacitors and cathodic sodium-, lithium-, and magnesium-ion battery applications\cite{zhang2016birnessite, alves2023,aranda2024improving}. In addition, its ion-exchange properties make it an effective material for water purification and heavy metal removal\cite{fanphotocatalytic}, while its strong catalytic activity allows it to play a crucial role in water splitting reactions for hydrogen (\ce{H2}) production\cite{lucht2015birnessite, peng2017redox}.

\begin{figure*}[htb]
    \centering
    \includegraphics[width=\linewidth]{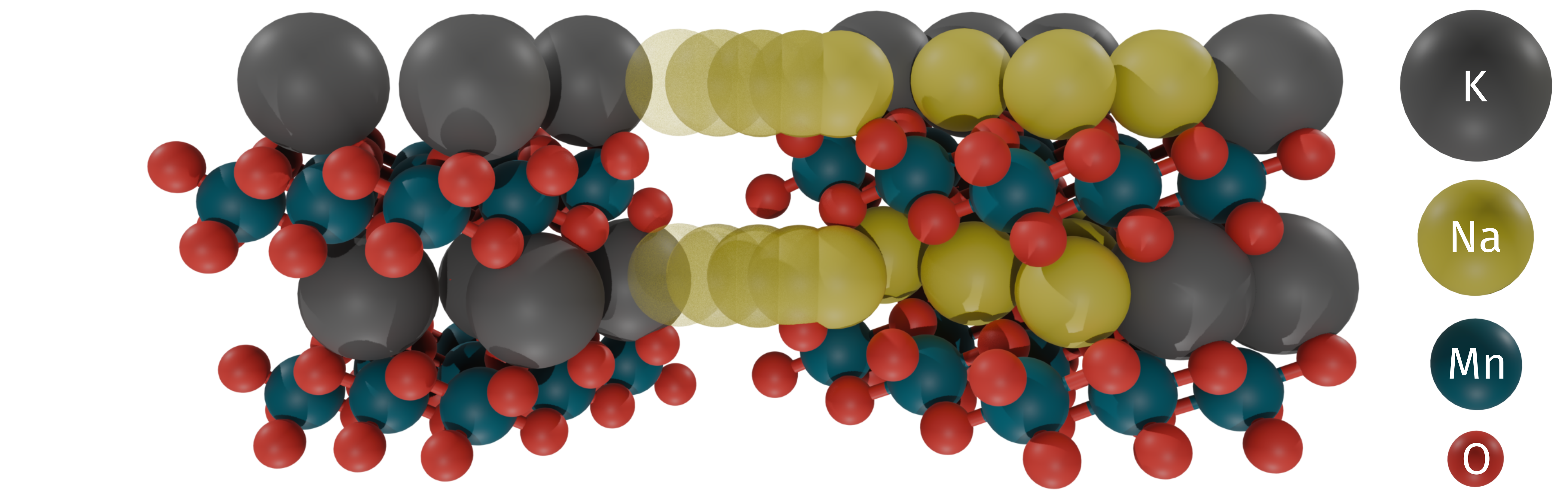}
    \caption{3D schematic depicting Sodium (\ce{Na+}, yellow atoms) intercalation in the Potassium (\ce{K+}, gray atoms) birnessite (KMO). The figure shows the KMO birnessite (left) and the \ce{Na+}-intercalated KMO structure (right).}
    \label{intercalation}
\end{figure*}

Understanding the theoretical properties and intercalation processes of $\delta$-\ce{MnO2} is a vital step to advance its potential innovative applications. From a computational perspective, studies have revealed the structure and redox properties\cite{peng2017redox}, the electronic structure in the presence of different cations\cite{lucht2015birnessite}, the defects properties of the \ce{MnO2} lattice\cite{kwon2009role, kwon2009zinc,kwon2013understanding}, the selectivity of different facets for the formation of reactive oxygen species\cite{yang2018insights}, the sorption selectivity\cite{kwon2013understanding,manceau2021nature}, among a multitude of other properties and applications of this material\cite{liu2018influence,manceau2022density,manceau2023density}.
Although these studies have revealed many of the important properties of $\delta$-\ce{MnO2}, there are still shortcomings in the current state of the computational literature. For example, although Raman spectroscopy is one of the main experimental techniques for characterizing birnessite, not many calculated Raman spectra have been presented. The available calculated spectra explain how water affects the spectra\cite{scheitenberger2021hidden}, however, they do not address the progressive inclusion of ionic intercalants such as \ce{Na+} and \ce{K+}, which are common in layered \ce{MnO2} structures. In addition, other properties of interest discussed in this work include formation energies, diffusion properties, structural features, and electronic structures during intercalation, which have yet to be fully described in the current state of the literature.

The present study constitutes a first-principles hybrid-level density functional theory (DFT) study of the \ce{Na+} ion intercalation process in a layered birnessite with the formula given by \ce{K_{1.33}Mn_3O_6}, \textit{i.e.}, $\mathrm{[K_{0.44}(Mn_{0.56}^{4+} Mn_{0.44}^{3+})O_2]}$, or simply denoted as KMO in this work. This particular formula is chosen to resemble the experimental high pressure synthesis of potassium birnessite\cite{chu2011buckled}. \ce{Na+} intercalation in the KMO structure is exemplified in \textbf{Figure \ref{intercalation}}. Furthermore, using a formation energy formalism, we analyze the stability of the structure as \ce{Na+} ions are intercalated between layers. Through this analysis, we propose a \ce{Na+} ion intercalation mechanism, which is further explored by analyzing the diffusion behavior of \ce{Na+} ions in the structure. Raman spectra simulations also provide information about the system's vibrational properties, where we find a relationship between the mode shifts and structural/symmetry properties induced by \ce{Na+} ion intercalation. Using X-ray diffraction (XRD) simulation and geometrical analysis, we describe the structure's evolution as \ce{Na+} ions are incorporated, resulting in modifications in the systems' lattice parameters and planar density. Finally, we describe the electronic properties of these systems \textit{via} spin-polarized densities of states.

This work improves our understanding of the \ce{Na+}-intercalation mechanism in potassium birnessite \ce{K_{1.33}Mn_3O_6}. We highlight energetic, structural, vibrational, and electronic adaptations along the intercalation process and discuss the limitations that it presents. This study aims to serve as a guide for experimental researchers performing electrochemical and structural analysis of KMO for pseudocapacitive energy storage, catalytic, and environmental purposes.

\section{Methodology}

Computational calculations were performed within the DFT framework using the CRYSTAL23 package\cite{erba2022crystal23}. Calculations were carried out defining the electronic exchange-correlation functional through the Heyd--Scuseria--Ernzerhof (HSE06) formulation. Unrestricted wave functions were considered due to the partially filled 3$d$ orbital electrons of the transition metal atoms in the structures. Grimme’s third-order dispersion correction, D3, with Becke--Johnson damping was utilized throughout all calculations to account for van der Waals and London dispersion forces\cite{grimme2010consistent,becke2005density}. Atomic orbitals were described using double-zeta valence with polarization (DZVP) quality basis sets \cite{vilela2019bsse}. XRD simulations were constructed using the VESTA visualization software\cite{momma2011vesta}. Raman simulations were achieved using the Coupled Perturbed Hartree-Fock/Kohn-Sham (CPHF/CPKS) methodology\cite{ferrero2008coupled, ferrero2008calculationofstat, ferrero2008calculationoffirst}. 

Integrations inside the first Brillouin zone were sampled on a Monkhorst-Pack $k$-mesh grid during both geometry optimizations and single-point energy calculations at a resolution of approximately 2${\pi}$/60 {\AA}${^{-1}}$ (${a \cdot k_a} =  40-60$, ${b \cdot k_b} =  40-60$, ${c \cdot k_c} =  40-60$, where \textit{a}, \textit{b}, and \textit{c} are direct lattice vectors and ${k_a}$, ${k_b}$, and ${k_c}$ are integer shrinking factors). To calculate the density matrix and Fermi energy, a denser “Gilat net” sampling of $k$-points was carried out with a resolution of approximately 2${\pi}$/120 \AA${^{-1}}$ (${k_G = 2k_{max}}$, where ${k_G}$ is the Gilat net shrinking factor and ${k_{max}}$ is the largest shrinking factor among the previously defined ${k_a}$, ${k_b}$, and ${k_c}$).

The convergence of the self-consistent field (SCF) energy, the root mean squared (RMS) forces, the maximum forces, the RMS atomic displacements, the maximum atomic displacements, and the convergence energies between geometries were established at 2.72$\times{10^{-6}}$ eV, 1.54$\times{10^{-2}}$ eV/\AA, 2.31$\times{10^{-2}}$ eV/\AA, 6.35$\times{10^{-4}}$ \AA, 9.53$\times{10^{-4}}$ \AA, and 2.72$\times{10^{-6}}$ eV, respectively. However, vibrational modes are highly sensitive to structural parameters, so we set a more rigorous convergence criterion for frequency calculations. The stricter thresholds for the convergence of the SCF energy, the RMS forces, the maximum forces, the RMS atomic displacements, the maximum atomic displacements, and the convergence energies between-geometry were established at 2.72$\times {10^{-10}}$ eV, 1.54$\times {10^{-3}}$ eV/{\AA}, 2.31$\times {10^{-3}}$ eV/\AA, 6.35$\times{10^{-5}}$ \AA, 9.53$\times{10^{-5}}$ \AA~and 2.72$\times{10^{-10}}$

The Nudged Elastic Band (NEB) method\cite{mills1995reversible,jonsson1998nudged} as implemented in the Vienna Ab initio Simulation Package (VASP)\cite{kresse1993ab,kresse1996efficient} was used to compute ion diffusion barriers. The Perdew--Burke--Ernzerhof (PBE)\cite{perdew1996generalized} functional was used during these calculations along with Grimme's D3 dispersion corrections\cite{grimme2010consistent}. The projector-augmented-wave (PAW) pseudopotentials (potpaw.64) were used \cite{blochl1994projector,kresse1999ultrasoft}. The cut-off energy during these calculations was 500 eV and the $k$-mesh was set to a $6\times6\times6$ Monkhost-Pack grid. For consistency with the rest of our work, the transition states obtained with this method were then exported to CRYSTAL23 and single-point energy calculations were performed with HSE06 using these geometries to obtain the final energy values. Note that minor errors might be present due to the use of optimized geometries at the PBE-level to produce HSE06 energies due to self-consistency.

\section{Results and Discussion}

\subsection{Geometry Optimization}\label{sec:GeomOpt}

\begin{figure*}[htb]
    \centering
    \includegraphics[width=0.95\linewidth]{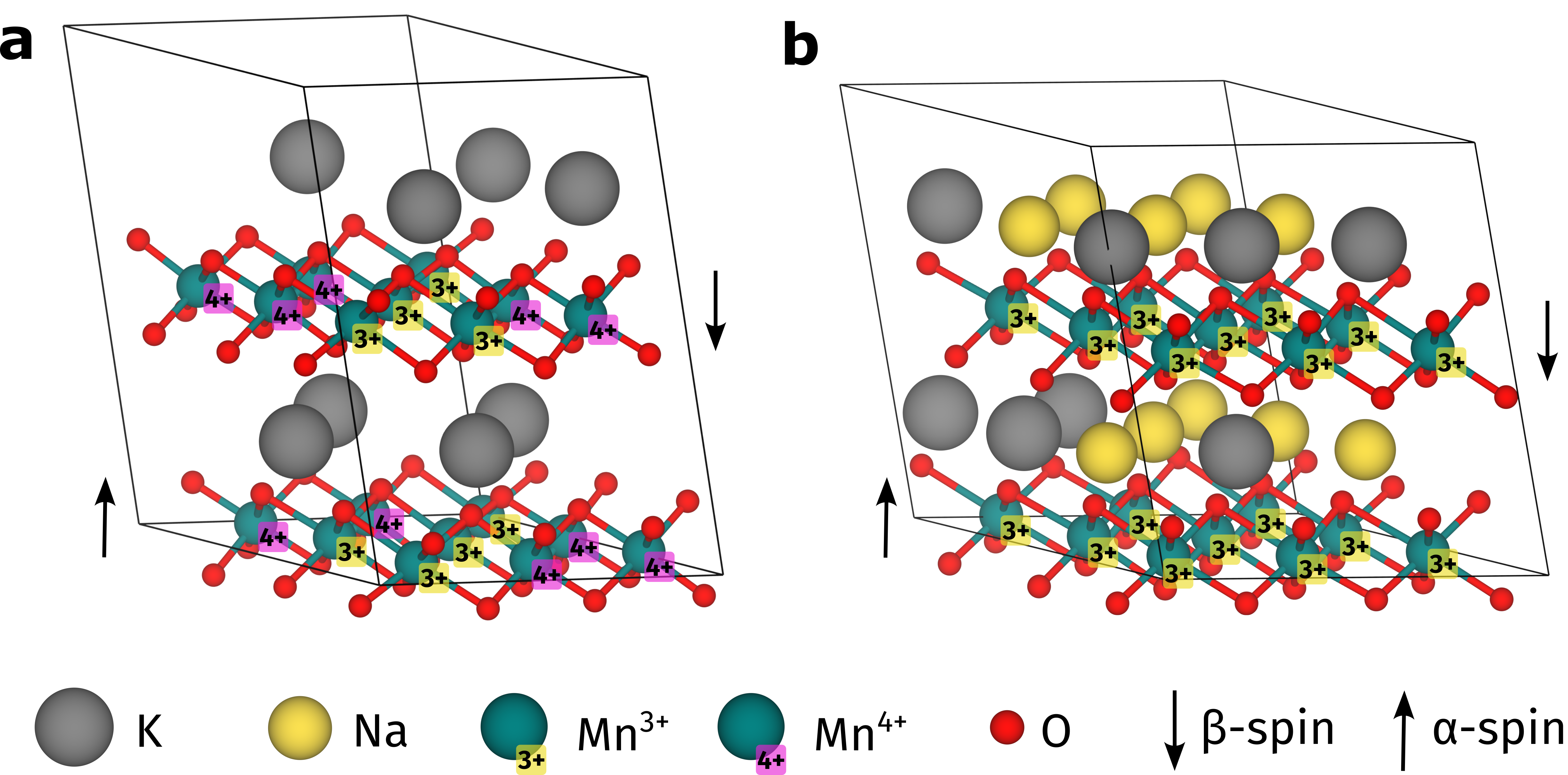}
    \caption{a) KMO birnessite with formula \ce{K8Mn18O36} and b) fully \ce{Na+}-intercalated birnessite with formula \ce{Na10K8Mn18O36}.}
    \label{Structures_Na0_Na10}
\end{figure*}

We generate a potassium birnessite host structure for sodiation with formula \ce{K8Mn18O36} (K$_{1.33}$Mn$_3$O$_6$) by systematically adding \ce{K+} to a $3\times3\times2$ supercell of $\delta$-\ce{MnO2}. \ce{K+} intercalations were tested \textit{via} single-point calculations on all available interlayer sites. For every additional \ce{K+}, only the most stable configuration is chosen. This verified the most stable positions for the \ce{K+} in the birnessite system within our hybrid DFT framework. The final structure at each intercalation step was then fully relaxed (atom positions and lattice parameters), and each relaxed structure served as an initial configuration for the next intercalation step. This process was repeated until the \ce{K8Mn18O36} stoichiometry was reached (shown in \textbf{Figure \ref{Structures_Na0_Na10}a}). The \ce{K+}-intercalation process generally results in a symmetry change for the $\delta$-MnO$_2$ system: from $P\bar{3}m1$ (without \ce{K+} ions) to $C2/\!m$ \cite{post2021raman, zhao2019construction} (fully intercalated with \ce{K+}). However, computationally, the symmetry was reduced to $P1$ to achieve the desired partially-intercalated \ce{K8Mn18O36} stoichiometry.

Throughout our calculations, the \ce{MnO2} layers were set to follow an out-of-plane antiferromagnetic order, typical of birnessite\cite{peng2017redox}, as shown in \textbf{Figure \ref{Structures_Na0_Na10}}. It is important to note that the partially-intercalated KMO structure was chosen as a starting point because of synthesis feasibility in an experimental setting. In particular, we based the host structure on the study by Chu \textit{et al.}\cite{chu2011buckled}, who investigated a potassium birnessite with the formula K$_{1.39}$Mn$_3$O$_6$. After obtaining the initial structure of K$_8$Mn$_{18}$O$_{36}$, insertion of \ce{Na+} ions followed, where a similar approach to \ce{K+}-intercalation was used until the saturation of the interlayer. Cell parameters for all \ce{Na+}-intercalated structures are shown in \textbf{Table S1}. The final stoichiometry of the fully co-intercalated \ce{Na+}/\ce{K+} birnessite was Na$_{10}$K$_8$Mn$_{18}$O$_{36}$ (Na$_{1.66}$K$_{1.33}$Mn$_3$O$_6$), as shown in \textbf{Figure \ref{Structures_Na0_Na10}b}. We note that the \ce{K+} and \ce{Na+} ions are initially added as neutral atoms and are ionized when added to the system due to charge compensation. Furthermore, the structures presented here represent an ideal system, where the layered structure is absent of defects, water, and external fields; therefore, some discrepancies with experimental analyzes are expected. 

Spin-difference ($\alpha - \beta$) Mulliken Population Analysis (MPA) was used to estimate the valence state of each Mn atom present in our KMO structures. It is important to note that even though MPA gives a qualitative indication of the transition metal valence states, it is known to yield inaccurate results for the number of unpaired electrons because of an overestimation of the bonds' covalent character. Thus, magnetic moment calculations would be necessary to accurately estimate the magnetic moment for each atom in the system. Nevertheless, by combining bond analysis and MPA, we were able to estimate the oxidation states. As shown in \textbf{Figure \ref{Structures_Na0_Na10}a}, the relationship Mn$^{3+}$/Mn$^{4+}$ matches the stoichiometry of the material, \textit{i.e.}, the number of Mn$^{3+}$ is equivalent to the amount of intercalated \ce{K+} ions. This relationship is consistent after complete interlayer saturation with Na$^+$ ions (\textbf{Figure \ref{Structures_Na0_Na10}b}). 

\subsection{Formation and binding energies}

\subsubsection{Formalism}

To understand the KMO sodiation process, we established a stepwise \ce{Na+}-insertion process determined by a defect formation energy (DFE) formalism based on chemical potential\cite{goyal2017computational} (see Section \ref{int_mechanism}). We select this method because our KMO structures are non-stoichiometric; therefore, they cannot be directly compared through total energies. Within this formalism, structures with varying amounts of interstitial atoms can be compared through their change in relative energy in a given range of chemical potential where crystal growth can occur\cite{maldonado2022atomic}. The formation energies were obtained using \textbf{Equation \ref{Formalism1}}:

\begin{equation}
    \Delta E_{F}(\mu)=[E_{D}-E_{H}]+\sum n_{i}\mu_{i}
    \label{Formalism1}
\end{equation}
Here, $\Delta E_{F}$ represents the resultant DFE, $E_{D}$ is the energy of the structures with defects (intercalants), $E_{H}$ the energy of the host material (all structures previously fully optimized), and $\mu_{i}$ is the chemical potential of the species $i$ involved in the formation of the material with defects.

Applying \textbf{Equation \ref{Formalism1}} to our KMO host material, with \ce{Na+}-intercalated KMO as defect structures, we obtained \textbf{Equation \ref{Formalism2}}:

\begin{eqnarray}
    \centering
    \Delta E_{F}(\mu)&=&[E_{D}-E_\mathrm{K_{8}Mn_{18}O_{36}}] +n_\mathrm{Mn}\mu_\mathrm{Mn}+n_\mathrm{O}\mu_\mathrm{O} + n_\mathrm{K}\mu_\mathrm{K}+n_\mathrm{Na}\mu_\mathrm{Na}
    \label{Formalism2}
\end{eqnarray}

Through \textbf{Equation \ref{Formalism2}}, we account for the variation in Mn, O and Na as independent variables and the resulting formation energy of the defect structures as dependent variables. Hence, to represent this equation in 3D, we take the variation of the chemical potential of Mn and O as a joint variable $\mu_\mathrm{Mn,O}$\, considered on the $x$-axis; Na variation, $\mu_\mathrm{Na}$\, on the $y$-axis; and the corresponding DFE, $\Delta E_{F}(\mu)$, on the $z$ axis. 

\subsubsection{Proposed sodium ion intercalation mechanism and stability of defect structures}
\label{int_mechanism}

To propose a \ce{Na+}-intercalation mechanism, we analyze potential sites for the introduction of this species to find preferential configurations. First, we model a \ce{Na}$_\mathrm{K}$ substitution. This structure resulted in a higher relative energy (by $1.36$--$4.18$ eV at different chemical potential ranges) than that of the host material. The increase in energy upon substitution occurs because the \ce{Na}$_\mathrm{K}$ energy accounts for both the removal of a \ce{K+} and the insertion of a \ce{Na+}. Therefore, substitutions were excluded for the rest of the analysis and the \ce{Na+} were incorporated into the KMO structure without removing any pre-intercalated \ce{K+}.

\begin{figure*}[hb!]
    \centering
    \includegraphics[width=\linewidth]{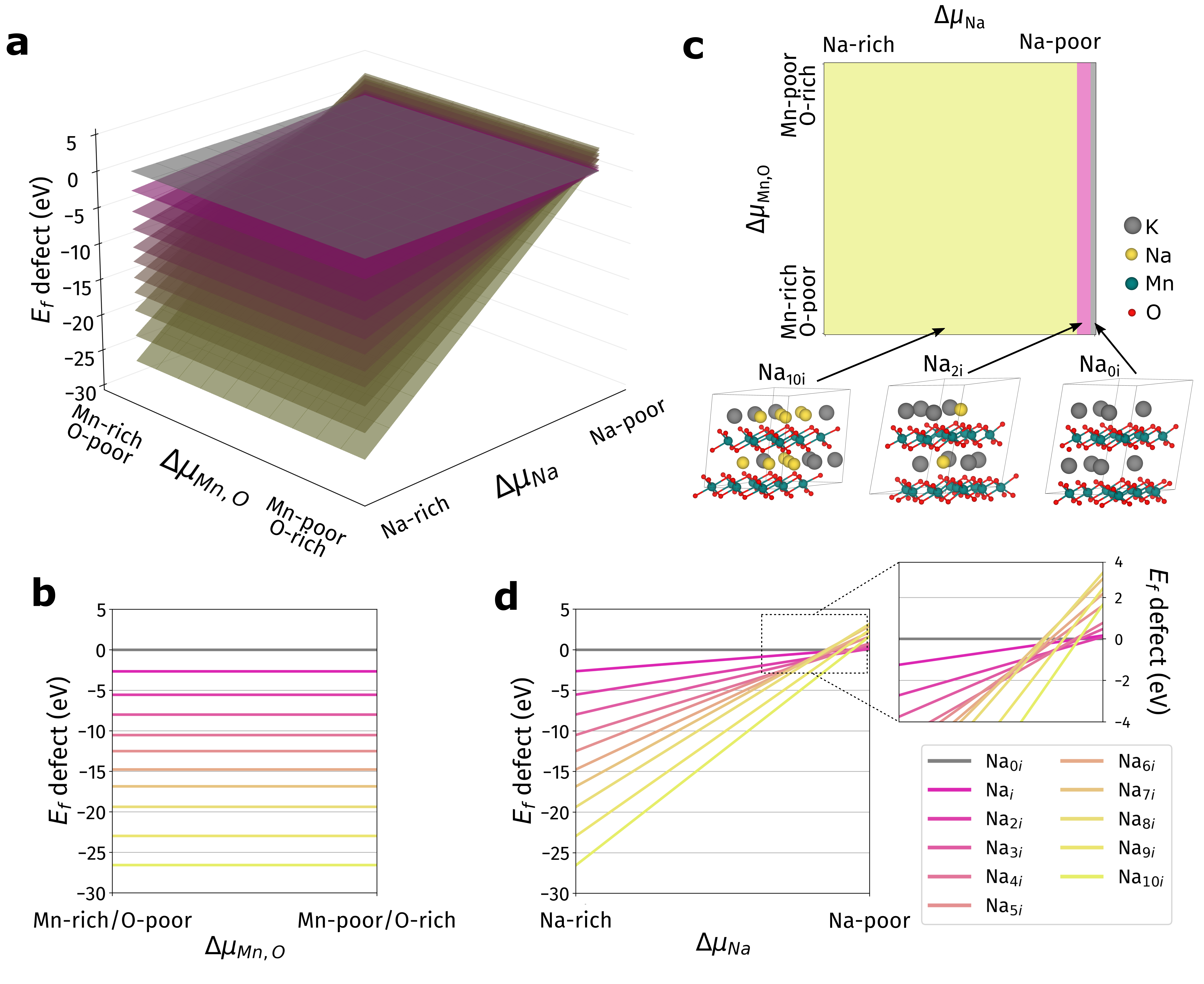}
    \caption{Defect formation energy (DFE) as \ce{Na+} ions are intercalated in the KMO birnessite. \textbf{a)} 3D DFE plot, considering the change in Mn, O, and Na growth conditions. \textbf{b)} Stability diagram (2D projection) showing the most stable structures. \textbf{c)} View along the Mn, O growth conditions, cross-section at \ce{Na+}-rich conditions. \textbf{d)} View along the Na growth conditions, cross-section at Mn-rich/O-poor conditions.}
    \label{DFE_all}
\end{figure*}

After systematically inserting \ce{Na+} into the KMO structure, we confirm the progressive reduction of the remaining Mn$^{4+}$ to Mn$^{3+}$. Once the available high-symmetry octahedral (hollow) lattice positions are full, the structure is considered saturated, as the addition of more \ce{Na+} would result in steric and Coulombic repulsions. This indicates that, without taking into account \ce{Na+}/\ce{K+} substitutions, the acceptance limit of the \ce{Na+} ions in the birnessite structure is equivalent to the Mn$^{4+}$ available in the rigid layers. In the case of our KMO structure, a maximum of 10 \ce{Na+} ions are accepted. 
Based on our stepwise \ce{Na+} insertion, we suggest that the barriers of ion insertion will determine the stability of the resulting structures.

\textbf{Figure {\ref{DFE_all}}a} shows the relative stability of the eleven optimized structures, from the initial KMO Na$_{0i}$ to Na$_{10i}$. Each plane represents the structures' stability at different chemical potentials. Here, a lower defect formation energy is associated with greater stability. The change in the chemical potential of Mn/O does not impact stability because their stoichiometry is unchanged from one structure to the next, as observed in \textbf{Figure {\ref{DFE_all}}b}. Rather, the stability is mediated by the sodium chemical potential, as the amount and position of sodium atoms change between the investigated structures. The 3D planes were projected onto the $x,y$  ($\Delta\mu_\mathrm{Mn,O}$,  $\Delta\mu_\mathrm{Na}$) plane, constructing a 2D defect stability diagram that exclusively shows the most stable structures and their chemical potential range (see \textbf{Figure \ref{DFE_all}c}). As shown, the fully saturated structure with 10 \ce{Na+}, is the most stable; even in regions near the \ce{Na+}-poor chemical potential limit. In the \ce{Na+}-poor limit, it can be observed that the most stable structures are Na$_{2i}$ and Na$_{0i}$. Finally, in \textbf{Figure {\ref{DFE_all}}d}, we show stability lines (plane projections on the $E_f$ vs. $\Delta \mu_\mathrm{Na}$ axes) with varying Na chemical potential. Under conditions rich in \ce{Na+}, the stability is directly correlated with the number of \ce{Na+} ions inserted into the material, i.e., more \ce{Na+} intercalants lead to greater stability. This trend is consistent through most of the chemical potential range. However, in near \ce{Na+}-poor conditions, the stability lines cross (see inset), allowing Na$_{2i}$ and Na$_{0i}$ to be the most stable in this range. Thus, the thermodynamic stability is governed by the chemical potential, except at the \ce{Na+}-poor limit.

\subsubsection{Binding energy}\label{sec:bindingenergy}

To further understand the energetics of the Na-intercalated KMO, we calculate the binding energy of our defect structures using the following equation: 

\begin{equation}
    \frac{E_B}{n} =-\frac{E_\mathrm{Structure}-E_\mathrm{Layer}-E_\mathrm{Ions}}{n}
    \label{eq_BE}
\end{equation}

\begin{figure*}[ht!]
    \centering
    \includegraphics[width=0.75\linewidth]{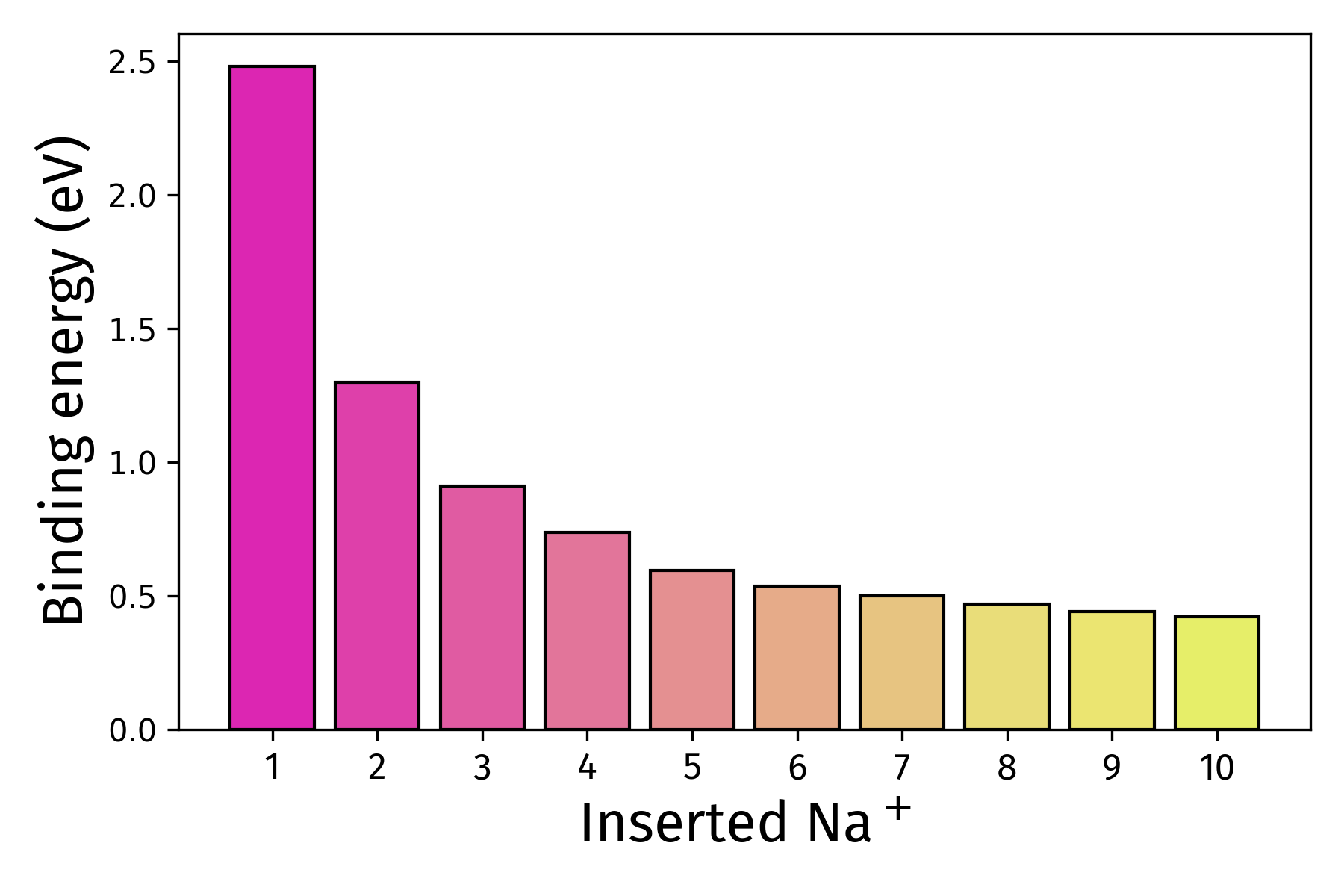}
    \caption{Binding energies as \ce{Na+} ions are intercalated in the KMO birnessite.}
    \label{BindingEnergy}
\end{figure*}

Here, $E_B/\!n$ is the binding energy per intercalated sodium ion. $E_\mathrm{Structure}$ is the energy of the complete structure with sodium intercalants. $E_\mathrm{Layer}$ is the energy of the structure without sodium intercalants (note that this energy includes pre-intercalated \ce{K+} and retains the geometry of $E_\mathrm{Structure}$ except for the removal of sodium). $E_\mathrm{Ions}$ is the energy of the \ce{Na+} ions by themselves without the KMO lattice and in the same positions as in $E_\mathrm{Structure}$. Finally, $n$ is the number of sodium intercalants in that structure. The results of the calculated binding energies are shown in \textbf{Figure \ref{BindingEnergy}}.

Binding energies can be interpreted as the energy required to separate a \ce{Na+} ion from the rest of the structure. Furthermore, because of the negative sign at the beginning of the equation, positive binding energies are favorable and indicate stronger interactions, whereas negative binding energies are unfavorable. 
In our particular structure, the introduction of a \ce{Na+} ion has a high binding energy of 2.5 eV and is strongly bound to the structure. However, a second \ce{Na+} intercalant is enough to lower the binding energy below 1.5 eV. The binding energies reduce as more \ce{Na+} is inserted into the structure. This reduction in binding energy is mainly due to (I) steric and Coulombic repulsions due to the presence of more interlayer ions, and (II) weaker electrostatic interactions due to the reduction of Mn$^{4+}$ to Mn$^{3+}$. The final structure (KMO-Na$_{10i}$) has a binding energy of 0.42 eV. Based on this analysis, we conclude that \ce{Na+} ions are easily removable in structures with a higher degree of saturation.
In other words, it is likely that, during the de-intercalation process, the first \ce{Na+} would be removed, while some sodium ions would remain in the interlayer.

\subsection{Diffusion Analysis}

\begin{figure*}[!htb]
    \centering
    \includegraphics[width=0.9\linewidth]{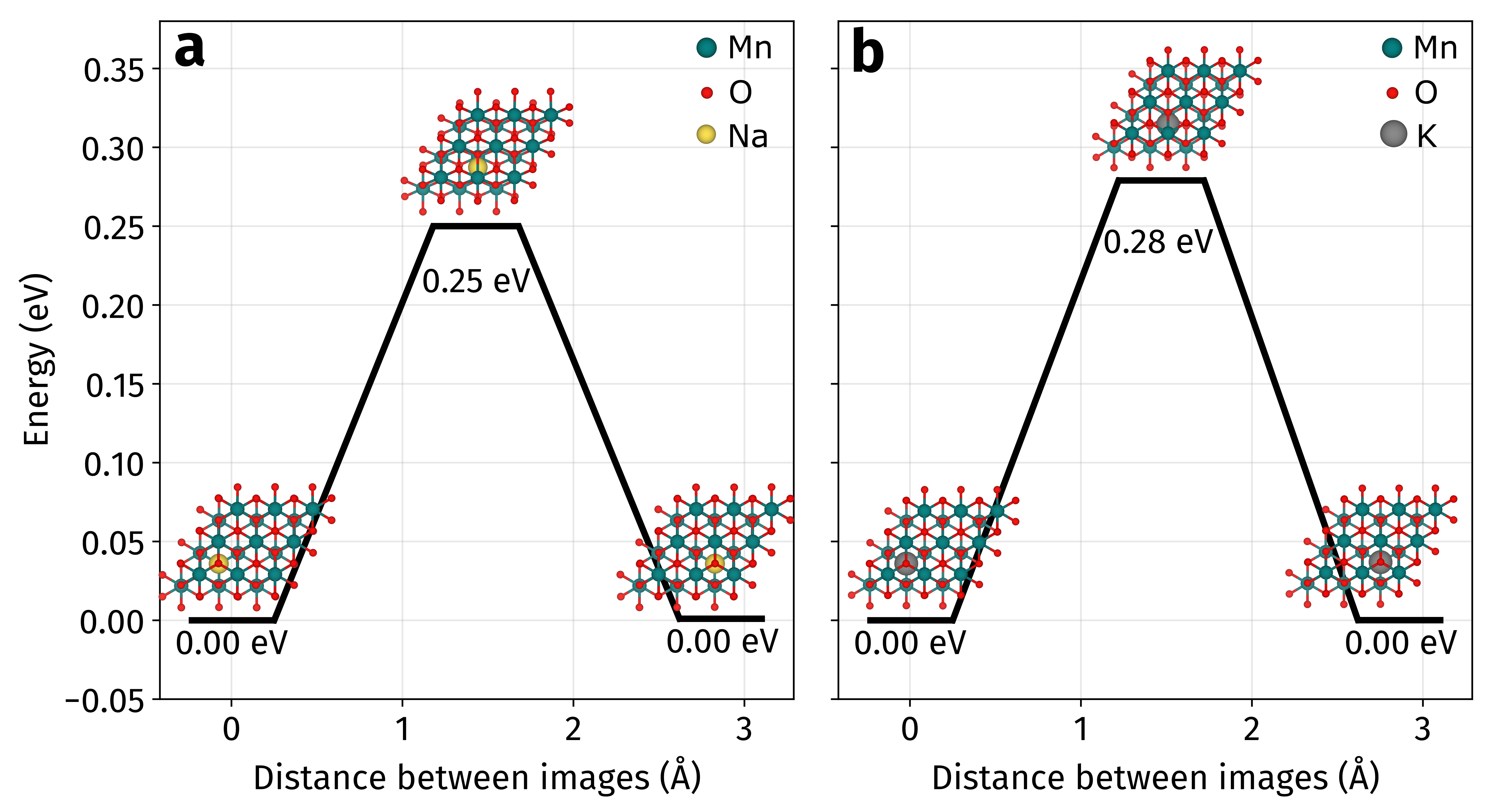}
    \caption{Diffusion analysis (activation barriers) calculated with NEB for a) \ce{Na+}, and b) \ce{K+} in the KMO structure.}
    \label{Diffusion}
\end{figure*}

 After analyzing the stability and binding energies of the \ce{Na+}-intercalation process, we examined ion diffusion at the interlayer by calculating the energy barriers present in the minimum energy pathway (MEP) to gain insight into the intercalation/deintercalation process. To achieve this, we performed Nudged Elastic Band (NEB) calculations. We are particularly interested in the difference between \ce{Na+}  \textit{vs.} \ce{K+} intercalation within the KMO structure, since ion dynamics are important descriptors of electrochemical processes. In this case, the calculations (see \textbf{Figure \ref{Diffusion}}) represent the movement of an ion from an octahedral (hollow) site, which is the lowest-energy position, to another adjacent octahedral site, passing through an intermediate position corresponding to the transition state. We note that the configuration chosen for this analysis is a simplified version, where all intercalants in the KMO structure are stripped out except for one. This simplified model is necessary because of the complexity of a partially intercalated structure, where there is likely a significant geometry rearrangement during intercalation, making it difficult to find the true lowest-energy and transition state positions.

The results of our simplified model (shown in \textbf{Figure \ref{Diffusion}}) demonstrate that the activation barrier for \ce{Na+} diffusion is lower than that of \ce{K+} by 0.03 eV ($\sim$10\%), indicating that \ce{K+} ions likely act as anchoring separators for the layers, while \ce{Na+} can preferentially diffuse in an electrochemical environment.

\subsection{Raman Spectroscopy Simulation}

\begin{figure*}[hbt!]
    \centering
    \includegraphics[width=0.9\linewidth]{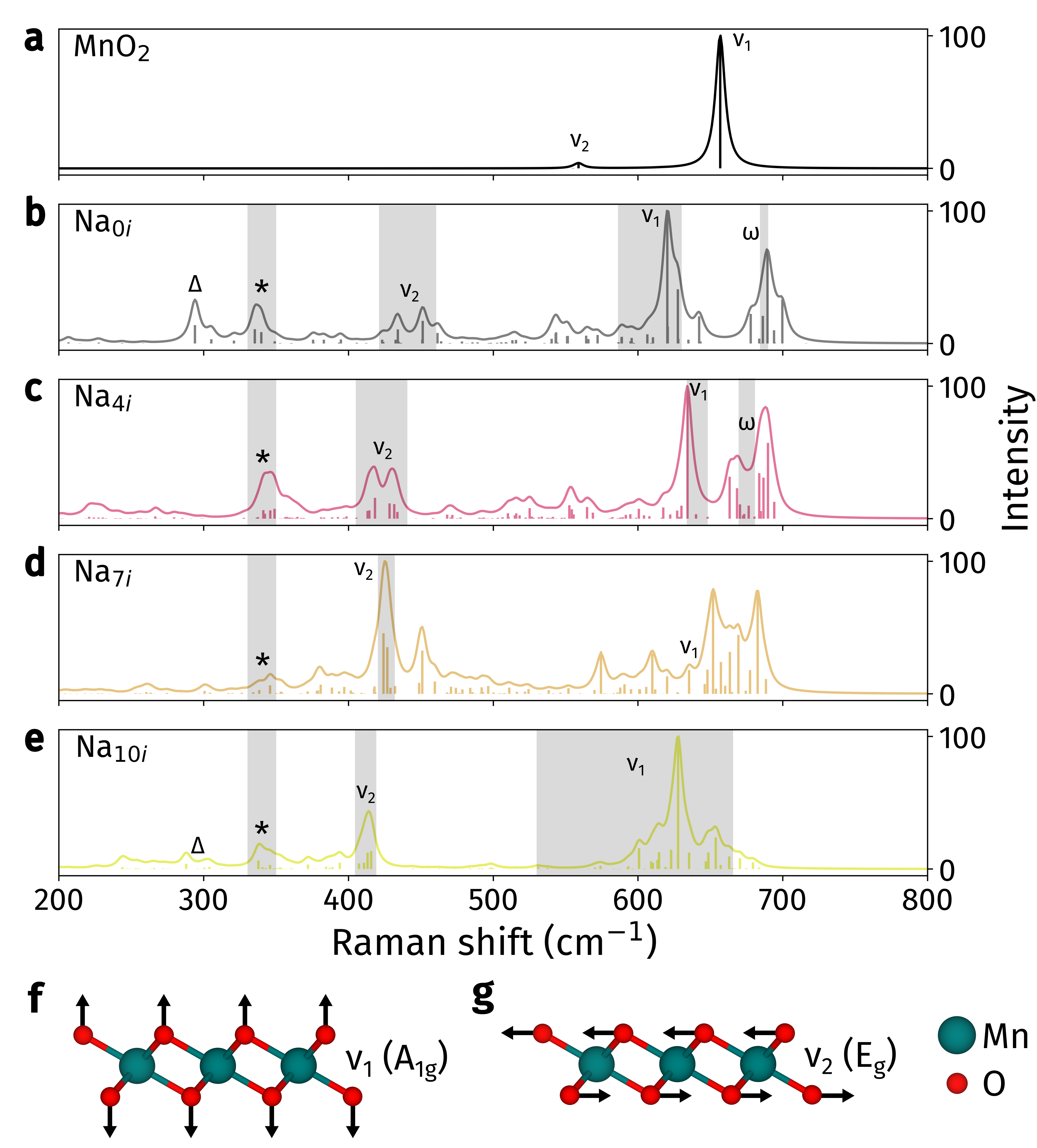}
    \caption{Simulated Raman Spectra using the HSE06 functional and CPHF/CPKS method for a) pristine $\delta$-\ce{MnO2}, b) potassium intercalated KMO $\ce{Na}_{0i}$, c) partially intercalated $\ce{Na}_{4i}$, d) partially intercalated $\ce{Na}_{7i}$, and e) fully intercalated $\ce{Na}_{10i}$ structures. The two characteristic birnessite modes are shown in f) $\nu_1$ (A\textsubscript{1g}) and g) $\nu_2$ (E\textsubscript{g}).}
    \label{Raman}
\end{figure*}

Raman spectra were simulated to further understand the characteristic vibrational modes of the birnessite KMO structure and the impact of \ce{Na+} introduced into the lattice (\textbf{Figure \ref{Raman}}). First, a pure $\delta-$\ce{MnO2} structure was simulated to determine the position and intensity of the main birnessite modes and compare them with the experimental literature results (\textbf{Figure \ref{Raman}a}). In experimental studies, two distinctive birnessite peaks are found. One mode occurs at 656 cm$^{-1}$, which can be viewed as the symmetric stretching vibration (Mn$-$O) of the \ce{MnO6} octahedral groups (A\textsubscript{1g} mode), and another at around 575 cm$^{-1}$, related to the stretching vibration (Mn$-$O) in the basal plane of [\ce{MnO2}] sheets (E\textsubscript{g} mode)\cite{julien2003raman}. Similarly, we found that only two signals appeared in the \ce{MnO2} spectrum: the A\textsubscript{1g} mode at 656 cm$^{-1}$ (labeled $\nu_1$ in \textbf{Figure \ref{Raman}}) and the E\textsubscript{g} mode at 558 cm$^{-1}$ (labeled $\nu_2$ in \textbf{Figure \ref{Raman}}). This close value from our quantum calculations to the experimental results validates the method used for our calculations.

As there are no previous reports of DFT simulated Raman spectra for our particular co-intercalated structure, we used this initial calculation as a set point for the progressive ion intercalation analysis of the following structures.

The KMO (Na$_{0i}$), Na$_{4i}$, Na$_{7i}$, and Na$_{10i}$ structures were selected due to their distinctive crystallographic structures, as identified from our XRD results (see Section \ref{XRD}), and were considered an appropriate representation of the gradual intercalation of Na$^+$. Consistent with experimental studies, a significant number of signals appear in the simulated spectra when ionic species are present in the interlayer\cite{julien2003raman,scheitenberger2021xrd}, compared to pure \ce{MnO2}. Structural symmetry, driven by sodium insertion, appears to play a major role in the activity of vibrational modes: the more symmetric the structure---both in the rigid lattice and the interlayer---the fewer active modes are observed.

When visualizing the atomic displacements for each computed mode, two general spectral regions were found across the four intercalated structures (\textbf{Figures \ref{Raman}b--e}): Region I, hosting modes associated with potassium/sodium ions at wavenumbers below 300--370 \ cm$^{-1}$; and Region II, hosting modes that correspond exclusively to the rigid lattice, at wavenumbers above 400 cm$^{-1}$. The A\textsubscript{1g} and E\textsubscript{g} modes (see \textbf{Figures \ref{Raman}f--g}) can be identified in region II. Although in the pure \ce{MnO2} spectrum these characteristic modes appear as sharp localized frequencies, the intercalated spectra exhibit near-symmetric vibrations with noticeable distortions and extend through a range of wavenumbers rather than at a specific value. We note that in Na$_{7i}$, these distortions hindered the recognition of characteristic vibrational modes. We attribute this to a pseudosymmetry transition that occurs at this intercalation step (see Section \ref{XRD}).

In Region I, the spectra display fewer signals, and their intensities decrease as sodiation takes place. All intercalated structures show a characteristic signal at around 330--350 cm$^{-1}$ (labeled \textbf{*} in \textbf{Figure \ref{Raman}}), seemingly related to the presence of ionic species in the interlayer. However, there was no clear symmetric vibration that could be associated with it. Another signal at 294--295 cm$^{-1}$ (labeled \textbf{$\Delta$} in \textbf{Figure \ref{Raman}}) can be clearly identified in the Na$_{0i}$ spectrum and is faintly visible in the Na$_{10i}$ spectrum; for this mode, intensity is the only observable difference between structures hosting exclusively \ce{K+} and those incorporating sodium.

We now focus on Region II. The A\textsubscript{1g} mode ($\nu_1$) initially redshifts from \ce{MnO2} to Na$_{0i}$, which is attributed to the symmetry breakdown of the \ce{MnO6} octahedra caused by Jahn-Teller distortion due to the presence of K$^{+}$\cite{ha2025raman}. Following this, $\nu_1$ generally shifts towards higher wavenumbers with progressive \ce{Na+} intercalation. In Na$_{0i}$, the mode spans 586--630 cm$^-1$, while in Na$_{4i}$ it narrows and increases to 634--648 cm$^{-1}$. In Na$_{7i}$, a slight trace is visible at 635 cm $^{-1}$, but its relatively low intensity reveals that it is not the dominant signal (which occurs at 651 cm $^{-1}$). Finally, in Na$_{10i}$, the mode extends broadly through 530--663 cm$^{-1}$, a range that we attribute to the enhanced pseudosymmetry achieved at full sodiation. The E\textsubscript{g} mode ($\nu_2$) also shows a strong initial redshift from \ce{MnO2} to Na$_{0i}$. Moreover, $\nu_2$ tends to shift to lower and more localized wavenumbers with increasing sodiation. In Na$_{0i}$, two distinct signals are observed that span 422--461 cm$^{-1}$, resulting from \ce{MnO6} symmetry breaking\cite{ha2025raman}. In Na$_{4i}$, the mode is displayed as vibrations located in layers between 405--438 cm$^{-1}$, with successive wavenumbers along this range activating vibrations in alternating layers. In Na$_{7i}$, the signals around 417--432 cm$^{-1}$ are consistent with the expected vibration, however, high spectral noise prevents a definitive assignment. Finally, in Na$_{10i}$, the mode is compressed into an individual well-resolved signal at 407--415 cm$^{-1}$ as the pseudosymmetry is regained. Pseudosymmetry recovery in Na$_{10i}$ also eliminated many of the non-totally symmetric (disorder-induced) modes that were present in partially-intercalated structures along the 450--550 cm$^{-1}$ range in Na$_{0i}$, Na$_{4i}$, and Na$_{7i}$.

The Na$_{0i}$ and Na$_{4i}$ spectra show an unidentified signal at wavenumbers above 660 cm$^{-1}$ (labeled $\omega$ in \textbf{Figure \ref{Raman}b--c}). In Na$_{0i}$, this signal displays a symmetric vibration at 685--690 cm$^{-1}$ seemingly along the basal plane, while in Na$_{4i}$ a similar vibration is observed around 670--680 cm$^{-1}$. In Na$_{7i}$, neither this symmetric movement nor any other pattern is clearly identified, although a signal is present. These signals appear to redshift with increasing sodiation, seemingly converging with the A\textsubscript{1g} mode as saturation is reached in Na$_{10i}$. 

Furthermore, along the spectra, several vibrational modes appear in pairs at adjacent wavenumbers. Each pair corresponds to the same atomic displacements, but occurs in alternating layers. For example, the E\textsubscript{g} mode ($\nu_2$)  in Na$_{10i}$ appears at 413 and 415 cm$^{-1}$ (\textbf{Figure \ref{Raman}e}). This splitting is attributed to subtle geometric and electronic variations between layers, arising from the nature and position of the ions between layers and the resulting Mn$^{3+}$/Mn$^{4+}$ ratio.

\subsection{XRD simulation and Geometry Analysis}\label{XRD}

As we modeled the systematic insertion of sodium ions in the KMO structure, we noticed a transformation in the overall unit cell of the structure, expressed as a deformation and an apparent increase in the crystalline order. To analyze this trend, we performed XRD simulations and geometry analysis.

\begin{figure*}[ht!]
    \centering
    \includegraphics[width=1\linewidth]{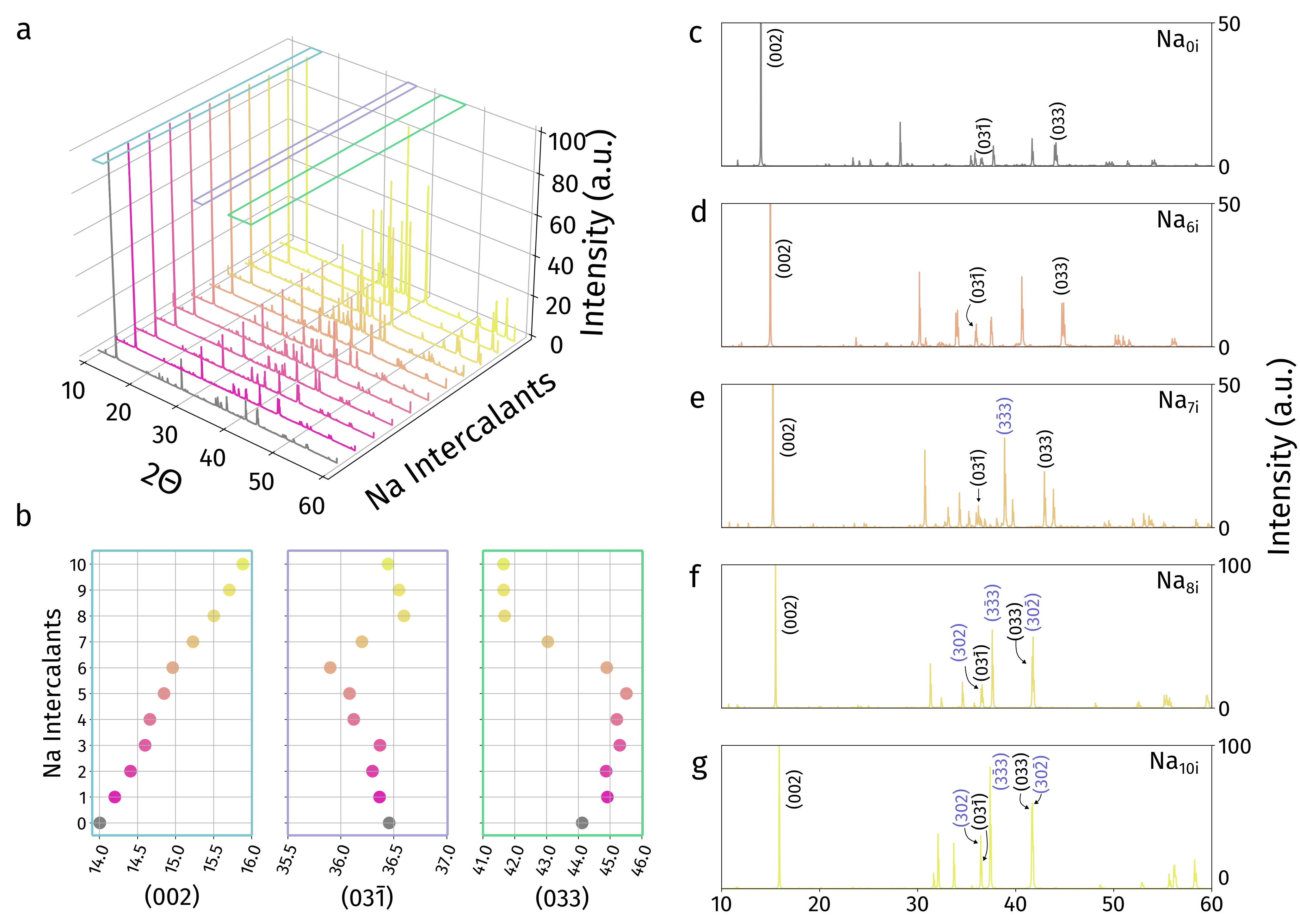}
    \caption{Simulated XRD patterns with increasing intercalated Na$^+$. \textbf{a)} 3D figure showing all of the simulated XRD patterns. Blue, lavender, and green rectangles show areas of interest for (002), (03$\bar{1}$), and (033) peaks, respectively. \textbf{b)} Projections along peaks of interest. \textbf{c-g)} Individual diffractograms for selected structures. Peaks that are maintained throughout the entire sodium intercalation process are labeled in black. Blue labels correspond to the largest peaks in the fully saturated structure (g). These peaks also appear in configurations Na$_{7i}$, Na$_{8i}$, and Na$_{9i}$.}
    \label{XRD_Na0_Na10}
\end{figure*}

\textbf{Figure \ref{XRD_Na0_Na10}a} shows a 3D visualization of the XRD patterns as sodium ions were intercalated in the structure. We first notice considerable peak shifts and intensity changes in the $2\theta$ range of 25 to 45$^\circ$, indicating plane distortion and the emergence of new diffracting planes, with an increase in intensity suggesting higher symmetry and atomic density.
The blue, purple and green rectangles show the areas of interest for the peaks corresponding to the planes (002), (03$\bar{1}$), and (033). The projection of these peaks is shown in \textbf{Figure \ref{XRD_Na0_Na10}b}.

The shift of the (002) peaks (leftmost plot in \textbf{Figure \ref{XRD_Na0_Na10}b}) to higher diffraction angles confirms a decrease in the interplanar distance, expressed as a distortion in the $c$ lattice parameters from 12.64 \,\AA~to 11.34 \,\AA. Although individual ions are more loosely bound to the structure as the interlayer is saturated (see Section \ref{sec:bindingenergy}), the decrease in interlayer distance suggests an increased force of attraction between the \ce{MnO2} and intercalated ion layers. Moreover, the (002) peak shows a change in its shift trend, starting with the configuration with seven Na$^+$ intercalants. We attribute this trend/slope change to a modification of the system's symmetry. This observation is further reinforced by changes in the shift slopes of the peaks (03$\bar{1}$) and (033) (center and right plots in \textbf{Figure \ref{XRD_Na0_Na10}b}), where we observe an even more drastic modification of the general trend of these lines, also starting with the Na$_{7i}$ configuration. 

The peak shifts shown in \textbf{Figure \ref{XRD_Na0_Na10}b} demonstrate a pseudo-symmetry recovery at the fully-intercalated limit. During the intercalation process, the introduction of potassium and sodium intercalants initially breaks down the pristine birnessite symmetry ($P\bar{3}m1$) to $P1$, as explained in 
Section \ref{sec:GeomOpt}. In the fully intercalated structure, the disorder introduced by the presence of two co-intercalated species preserves $P1$ symmetry; however, the underlying structure (i.e. if all intercalants were the same) has $C2/\!m$ symmetry. 

Next, we analyze the lengths of the Mn-O bonds during the stepwise reductions of Mn$^{4+}$/Mn$^{3+}$, as shown in Table \ref{MnO bonds table}. With each Mn$^{4+}$/Mn$^{3+}$ reduction, the average Mn-O bond length increases due to symmetry breaking, which confirms the Jahn–Teller distortion of the MnO${_6}$ octahedra. This distortion also explains the expansion of the layers along the [\={1}10] direction, as evidenced by the increase in the lattice parameters $a$ and $b$.

As the lattice parameters change, the planes (3\={3}4), (30\={3}), (3\={3}0) and (301) are converted into the planes (3\={3}3), (30\={2}), (3\={3}\={1}), and (302), respectively; all of these planes intersect Mn, O, and interlayer atoms. The increased intensities of these planes, together with that of the (400) plane, indicate improved atomic accommodation and higher symmetry, due to the higher planar density of the interlayer.

\begin{table*}
\centering
\caption{Mn--O bond-length variation as a consequence of Mn$^{4+}$/Mn$^{3+}$ reduction.}
\vspace{2mm}
  \begin{tabular}{cccc}
    \hline
    Structure & Mn$^{4+}$/Mn$^{3+}$ & Mn--O bond- & Variance of\\
    & ratio &  length average  (\AA) &  Mn-O bond-length (\%)\\
    \hline
    & & \multirow{3}{*}{1.951} &  \\
    KMO (Na$_\mathrm{{0i}}$) & 10/8  & & 1.09\\
    &  & & \\
    \hline
    & & \multirow{3}{*}{2.009} &  \\
    Na$_\mathrm{{5i}}$ & 5/13  & & 2.46\\
    &  & & \\
    \hline
    & & \multirow{3}{*}{2.084} &  \\
    Na$_\mathrm{10i}$ & 0/18  & & 4.79\\
    & & & \\
    \hline
  \end{tabular}
\label{MnO bonds table}
\end{table*}

\subsection{Electronic Density of States (DOS)}
\begin{figure*}[hbt!]
    \centering
    \includegraphics[width=\linewidth]{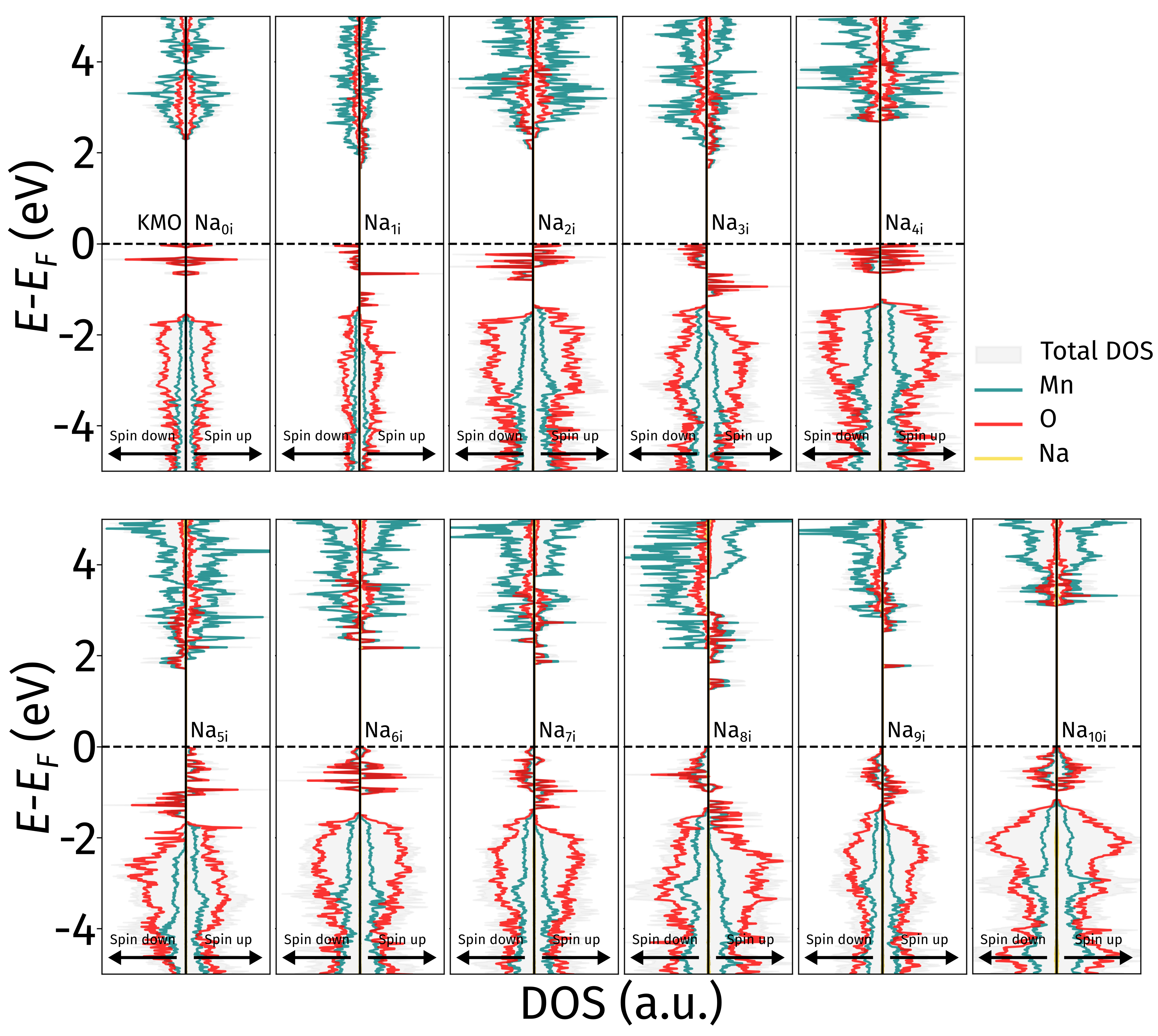}
    \caption{Electronic density of states for the pristine and intercalated KMO birnessite structures as Na$^+$ ions are added to the lattice.}
    \label{DOS}
\end{figure*}

The spin-polarized, atom-projected densities of states (DOS) of all KMO structures are plotted and compared in \textbf{Figure \ref{DOS}} (see \textbf{Figure S1} for Mn \textit{d}-orbital contributions). The experimental pristine birnessite ($\delta-$\ce{MnO2}) has a reported band gap of 2.23 eV, and our pristine KMO model (Na$_{0i}$) shows excellent agreement with a band gap of 2.34 eV. The fully-saturated structure (Na$_{10i}$) presents a band gap of 3.11 eV, and intermediate structures have varied band gap values. 

All structures exhibit either semiconducting or insulating behavior. Although Na$^+$ ions do not appear to be prevalent contributors to DOS around the Fermi level, it is evident that there are electronic effects that arise as a result of the introduction of sodium into the structure.The difference in electronic properties arises primarily due to the modification of the Mn$^{4+}$/Mn$^{3+}$ oxidation states as \ce{Na+} is intercalated, which in turn causes lattice distortions (previously analyzed in \textbf{Table \ref{MnO bonds table}}) and symmetry effects (as analyzed by XRD). 

The interplay between rigid \ce{MnO2} layers and \ce{Na+} intercalants results in a distinct electronic pattern. Structurally, the \ce{Na+} ions intercalate in opposite interlayers. This results in a reduction of Mn$^{4+}$ to Mn$^{3+}$ in opposite rigid layers. Due to the initial antiferromagnetic spin alignment, a decompensation/compensation effect is observed as Mn is reduced, leading to distinct spin-polarized states. In particular, when a \ce{Na+} is introduced, the reduced Mn$^{3+}$ sites have one extra electron, which aligns with the layer and decompensates the antiferromagnetic arrangement. This is in agreement with previous studies\cite{PengH2017Redox} showing that K$^+$ intercalants induce small polarons in birnessite, resulting in energy-split conduction band edges between the two spin-polarized states. Similarly, we observe a spin-polarized effect in the DOS around the Fermi level (CBM and VBM), which is particularly strong for odd intercalation numbers, since these have one extra unpaired electron in the layers. The exception being Na$_{8i}$, since the ground state for this structure resulted in two uncompensated Mn$^{3+}$ atoms in a single layer. This results in spin-down polarized states directly below the Fermi level and spin-up polarized states directly above it.

We note that the structures Na$_{i}$, Na$_{3i}$, Na$_{5i}$, Na$_{7i}$, Na$_{8i}$, and Na$_{9i}$ all present a spintronic DOS profile resembling that of bipolar magnetic semiconductors. These materials have spin-polarized VBM and CBM in opposing spin directions\cite{li2016first}. By controlling the position of the Fermi level, for example via a gate voltage, it is possible to access fully spin-polarized conducting states at the valence/conduction band. This opens the possibility of using co-intercalated birnessite as a spin filter with bipolar field effect, field effect spin valve, or to detect and separate entangled electrons from superconductors\cite{li2013bipolar}.

\textbf{Table S2} shows the band gap for all Na$^+$-intercalated structures. The band gaps span a range of values, from 1.94 to 3.13 eV; thus, it is possible to systematically fine-tune the band gap depending on the number of intercalants. The minimum band gap (1.94 eV) occurs at 5 Na$^+$ intercalants in the alpha channel (spin-up). In general, structures with strong spin polarization (odd intercalation number and Na$_{8i}$) are observed to have lower band gaps (in one of their spin channels) than structures with fully compensated antiferromagnetic oxidation/spin patterns. The maximum band gap within our structures (3.13 eV) is reached when the intercalation sites are completely saturated. Here, the structure returns to being fully spin compensated, with an even distribution of spin-up and spin-down electrons, where additionally all \ce{MnO6} octahedra have Jahn-Teller distortions. 

\section{Conclusion}
In conclusion, we provide a systematic analysis of the structural, energetic, diffusional, vibrational (Raman), and electronic/spintronic properties of  a sodium-intercalated potassium birnessite. We begin with a systematic addition of \ce{K+} to $\delta$-\ce{MnO2} birnessite to find the ground state of a potassium birnessite structure with \ce{K_{1.33}Mn3O6} stoichiometry. This particular stoichiometry was chosen due to the previously-reported experimental synthesis feasibility. We take this as our starting point and incorporate sodium until we reach a stoichiometry of \ce{Na_{1.66}K_{1.33}Mn3O6}, with fully intercalated layers. Our findings include the following: (1) The most stable structures are the fully intercalated Na$_{10i}$, partially intercalated Na$_{2i}$ and Na$_{0i}$ structures, in different regions of the \ce{Na} chemical potential. (2) Binding energies indicate that Na$^+$ ions are more loosely bound to the lattice, especially near the saturation limit. (3) The diffusion of Na${^+}$ is more energetically favorable than that of K${^+}$, allowing \ce{K+} ions to serve as spacers, while Na$^+$ diffuses quickly through the structure. (4) Characteristic Raman peaks are shifted when sodium and potassium are added to the structure. In addition, the number of peaks is reduced as the structure regains symmetrical features when fully saturated. (5) The XRD patterns show a change in symmetric order at Na$_{7i}$, indicated by a change in the trend of the characteristic peaks. Furthermore, at Na$_{10i}$, we observe a pseudosymmetry recovery when the interlayer is fully saturated. (6) DOS plots reveal that the band gaps and magnetic behavior of the system can be controlled by intercalation. Some of the spin-polarized structures behave as bipolar magnetic semiconductors, with potential application in spintronics devices. Furthermore, band gaps seem to depend on the number of intercalants. This work provides an overarching analysis of intercalated birnessite and can serve as an enriching complement to current and future experimental and theoretical studies.

\bibliographystyle{ieeetr}
\bibliography{ref}

\newpage

\title{Supplementary Information for ``How Does Intercalation Reshape Layered Structures? A First-Principles Study of Sodium Insertion in Layered Potassium Birnessite''}

\date{}

\renewcommand{\thesection}{S\arabic{section}}
\renewcommand{\thepage}{S\arabic{page}}  
\renewcommand{\thetable}{S\arabic{table}}
\renewcommand{\thefigure}{S\arabic{figure}}

\setcounter{page}{1}
\setcounter{section}{0}
\setcounter{figure}{0}
\setcounter{table}{0}

\maketitle

\begin{table*}
\centering
\caption{Crystal unit cell parameters}
\vspace{2mm}
\begin{tabular}{ccccccc}
Structure & a (\AA)  & b (\AA)     & c (\AA)     & $\alpha$ ($^{\circ}$)  & $\beta$ ($^{\circ}$)  & $\gamma$ ($^{\circ}$)  \\ \hline
&&\\[-3mm]
Na$_{1i}$ & 8.84  & 8.84 & 12.76 & 101.13 & 79.23 & 122.81   \\\cline{0-6}
&&\\[-3mm]
Na$_{2i}$ & 8.88  & 8.89 & 12.57 & 101.00 & 79.40 & 123.11   \\\cline{0-6}
&&\\[-3mm]
Na$_{3i}$ & 9.00  & 8.96 & 12.43 & 101.58 & 79.09 & 123.72   \\\cline{0-6}
&&\\[-3mm]
Na$_{4i}$ & 9.04  & 9.04 & 12.43 & 102.15 & 77.97 & 124.08   \\\cline{0-6}
&&\\[-3mm]
Na$_{5i}$ & 9.10 & 9.03 & 12.21 & 101.60 & 79.90 & 124.03   \\\cline{0-6}
&&\\[-3mm]
Na$_{6i}$ & 9.15  & 9.16 & 12.11 & 101.00 & 79.30 & 124.64   \\\cline{0-6}
&&\\[-3mm]
Na$_{7i}$ & 9.25  & 9.27 & 11.66 & 93.19 & 85.59 & 125.00   \\\cline{0-6}
&&\\[-3mm]
Na$_{8i}$ & 9.37 & 9.36 & 11.63 & 93.60 & 79.66 & 125.38   \\\cline{0-6}
&&\\[-3mm]
Na$_{9i}$ & 9.47  & 9.46 & 11.46 & 93.19 & 80.15 & 125.95  \\\cline{0-6}
&&\\[-3mm]
Na$_{10i}$ & 9.57  & 9.57 & 11.34 & 93.23 & 80.13 & 126.53   \\\cline{0-6}
\end{tabular}
\label{CellParameters}
\end{table*}

\begin{table*}
\centering
\caption{Band gaps of the KMO and all Na$^+$ intercalated structures.}
\vspace{2mm}
\begin{tabular}{|ccc|}
\hline
\multicolumn{1}{|c|}{\textbf{Intercalant}} & \multicolumn{1}{c|}{\textbf{Alpha band gap (eV)}} & \textbf{Beta band gap (eV)} \\ \hline
\textbf{
KMO}                          & 2.31                                            & 2.31                      \\
\textbf{Na$_{1i}$}                               & 2.31                                            & 1.98                      \\
\textbf{Na$_{2i}$}                               & 2.36                                            & 2.32                      \\
\textbf{Na$_{3i}$}                               & 2.34                                            & 2.17                      \\
\textbf{Na$_{4i}$}                               & 2.75                                            & 2.81                      \\
\textbf{Na$_{5i}$}                               & 1.95                                            & 2.73                      \\
\textbf{Na$_{6i}$}                               & 2.17                                            & 2.39                      \\
\textbf{Na$_{7i}$}                               & 2.06                                            & 2.33                      \\
\textbf{Na$_{8i}$}                               & 2.09                                            & 2.19                      \\
\textbf{Na$_{9i}$}                               & 2.18                                            & 2.87                      \\
\textbf{Na$_{10i}$}                              & 3.13                                            & 3.11                      \\ \hline
\end{tabular}

\label{Bandgaps}
\end{table*}

\begin{figure}
    \centering
    \includegraphics[width=\linewidth]{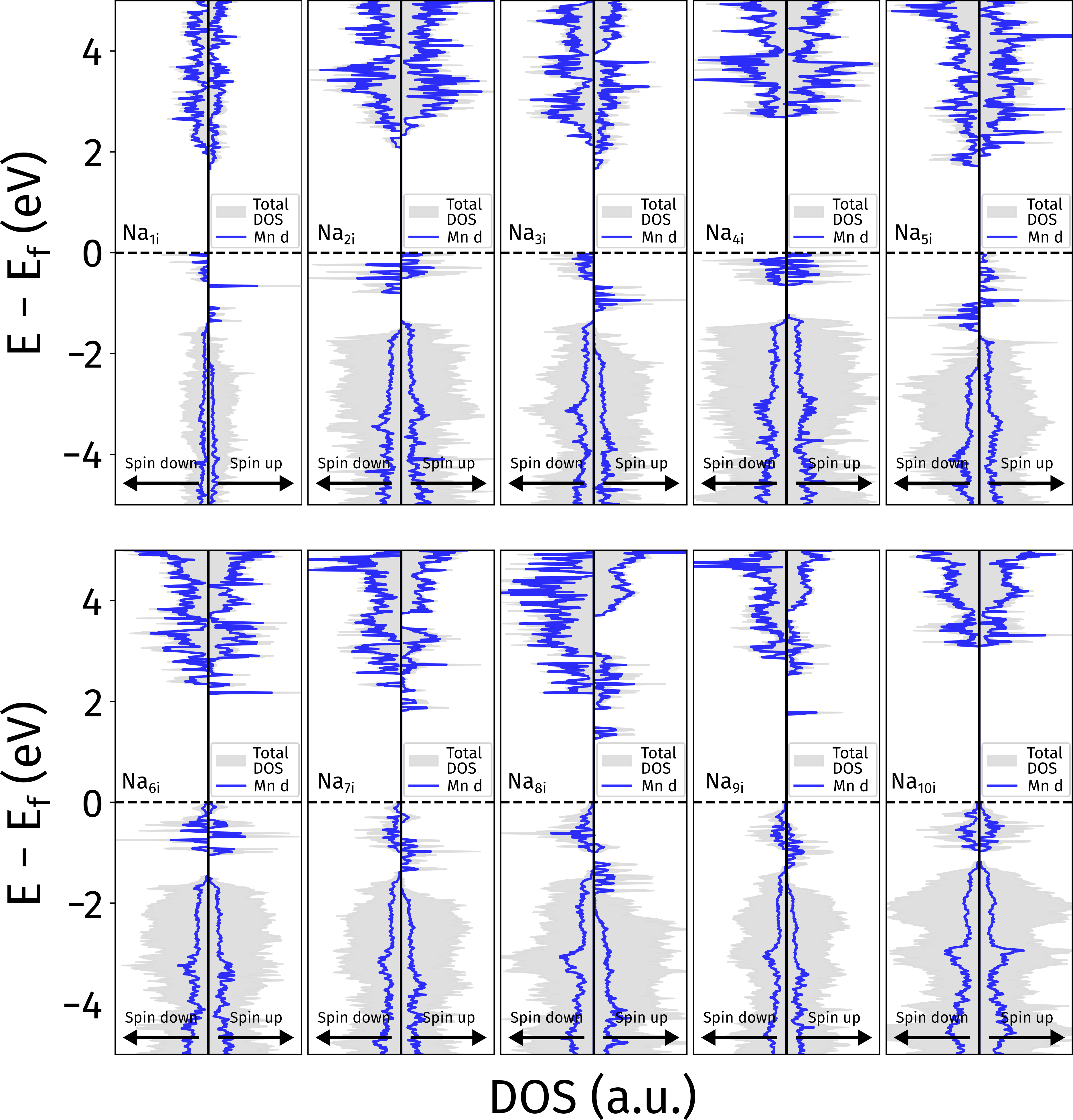}
    \caption{$d$-orbital projection for all intercalated KMO structures with Na$^+$.}
    \label{fig:DOSMn-d}
\end{figure}

\newpage

\section*{CIF files}

KMO structure (Na$_{0i}$)
\footnotesize
\begin{verbatim}
_cell_length_a                         8.786108
_cell_length_b                         8.804715
_cell_length_c                         12.825908
_cell_angle_alpha                      98.813900
_cell_angle_beta                       81.509062
_cell_angle_gamma                      122.452878
_symmetry_space_group_name_H-M         'P 1'
_symmetry_Int_Tables_number            1

loop_
_symmetry_equiv_pos_as_xyz
   'x, y, z'

loop_
   _atom_site_label
   _atom_site_type_symbol
   _atom_site_fract_x
   _atom_site_fract_y
   _atom_site_fract_z
 K001   K   0.157879   0.421706   0.242601
 K002   K  -0.433456  -0.137844   0.244675
 K003   K   0.111401  -0.189641  -0.245435
 K004   K  -0.346734  -0.075828  -0.256593
 K005   K   0.057465   0.359445  -0.254909
 K006   K  -0.399931   0.303673   0.255712
 K007   K  -0.484842   0.244204  -0.242919
 K008   K   0.006687  -0.254748   0.257108
Mn009  Mn   0.498562  -0.175841   0.493854
Mn010  Mn  -0.166235   0.489763  -0.496854
Mn011  Mn   0.497481   0.155148  -0.497141
Mn012  Mn   0.165760   0.494609   0.492074
Mn013  Mn  -0.002173   0.339639  -0.003428
Mn014  Mn   0.335826   0.006868   0.006315
Mn015  Mn   0.000385  -0.327915   0.006626
Mn016  Mn  -0.331065   0.011338  -0.003154
Mn017  Mn  -0.169322   0.159212   0.493676
Mn018  Mn   0.168812  -0.173069  -0.497259
Mn019  Mn   0.332590  -0.323895  -0.002638
Mn020  Mn  -0.331549   0.341904   0.007437
Mn021  Mn   0.003652   0.005165  -0.005058
Mn022  Mn  -0.498399   0.492191  -0.499402
Mn023  Mn   0.336236   0.343414  -0.002426
Mn024  Mn  -0.334348  -0.324254  -0.001064
Mn025  Mn  -0.168682  -0.176846  -0.496214
Mn026  Mn   0.163599   0.161825  -0.494545
 O027   O   0.384080  -0.061026   0.416189
 O028   O  -0.064907   0.392410  -0.425607
 O029   O  -0.399057   0.059879  -0.424559
 O030   O   0.064123  -0.404894   0.419603
 O031   O  -0.099709   0.442213  -0.073826
 O032   O   0.438446  -0.090544   0.079105
 O033   O   0.101708  -0.428852   0.079834
 O034   O  -0.434494   0.106718  -0.075210
 O035   O  -0.272064   0.256680   0.420907
 O036   O   0.265611  -0.276195  -0.426628
 O037   O   0.231597  -0.226108  -0.074717
 O038   O  -0.217434   0.227388   0.083623
 O039   O   0.113756  -0.102744   0.084743
 O040   O  -0.113044   0.120415  -0.090104
 O041   O  -0.385439   0.380941  -0.417320
 O042   O   0.383319  -0.403604   0.408873
 O043   O   0.437212   0.228953   0.085129
 O044   O   0.222695   0.455197  -0.083603
 O045   O  -0.448287  -0.214798  -0.082649
 O046   O  -0.216587  -0.429242   0.090687
 O047   O  -0.055911  -0.288949  -0.416294
 O048   O  -0.270899  -0.063292   0.415571
 O049   O   0.052260   0.268717   0.414723
 O050   O   0.279350   0.045883  -0.410260
 O051   O   0.241678   0.094684  -0.077597
 O052   O   0.098907   0.240851   0.065408
 O053   O  -0.231750  -0.086540   0.068122
 O054   O  -0.093024  -0.235253  -0.072236
 O055   O  -0.259802  -0.419278   0.423099
 O056   O  -0.393487  -0.283616  -0.422384
 O057   O   0.259983   0.402613  -0.428614
 O058   O   0.398531   0.253459   0.431746
 O059   O  -0.439569   0.450378  -0.076736
 O060   O   0.426305  -0.414169   0.077335
 O061   O   0.067495  -0.074240   0.434658
 O062   O  -0.074526   0.072406  -0.422051
\end{verbatim}
\normalsize

KMO structure with one Na$^+$ intercalant (Na$_{1i}$)
\footnotesize
\begin{verbatim}
_cell_length_a                         8.844811
_cell_length_b                         8.844012
_cell_length_c                         12.763156
_cell_angle_alpha                      101.130567
_cell_angle_beta                       79.279611
_cell_angle_gamma                      122.812056
_symmetry_space_group_name_H-M         'P 1'
_symmetry_Int_Tables_number            1

loop_
_symmetry_equiv_pos_as_xyz
   'x, y, z'

loop_
   _atom_site_label
   _atom_site_type_symbol
   _atom_site_fract_x
   _atom_site_fract_y
   _atom_site_fract_z
 K001   K  -0.003122   0.362387   0.248945
 K002   K  -0.450076  -0.209793   0.244507
 K003   K   0.154735  -0.138548  -0.248199
 K004   K  -0.358559  -0.024173  -0.257665
 K005   K   0.041822   0.379546  -0.260736
 K006   K  -0.459636   0.146145   0.251514
 K007   K  -0.494144  -0.498446  -0.251265
 K008   K  -0.033696  -0.261006   0.263327
MN009  MN  -0.483964  -0.176143   0.488093
MN010  MN  -0.148551   0.496464   0.499818
MN011  MN  -0.484683   0.158525  -0.498580
MN012  MN   0.184338   0.494465   0.484835
MN013  MN  -0.001466   0.339978  -0.005314
MN014  MN   0.334840   0.008736   0.014195
MN015  MN   0.000394  -0.327060   0.009056
MN016  MN  -0.330247   0.010954  -0.002874
MN017  MN  -0.151607   0.159895   0.485122
MN018  MN   0.183398  -0.170825   0.496623
MN019  MN   0.331811  -0.323821  -0.005483
MN020  MN  -0.334338   0.344630   0.013866
MN021  MN   0.003885   0.005453   0.000783
MN022  MN  -0.483452   0.495249  -0.495949
MN023  MN   0.335518   0.342366  -0.002654
MN024  MN  -0.333433  -0.326769  -0.004795
MN025  MN  -0.154312  -0.170343  -0.488203
MN026  MN   0.181648   0.165776  -0.488614
 O027   O   0.404174  -0.061408   0.409446
 O028   O  -0.050405   0.395845  -0.435721
 O029   O  -0.382829   0.064404  -0.433144
 O030   O   0.070128  -0.390551   0.408346
 O031   O  -0.098931   0.440466  -0.070523
 O032   O   0.435987  -0.093003   0.081748
 O033   O   0.100661  -0.425457   0.079789
 O034   O  -0.434075   0.108502  -0.066143
 O035   O  -0.252003   0.257783   0.413013
 O036   O   0.281367  -0.268987  -0.433901
 O037   O   0.234938  -0.223635  -0.072823
 O038   O  -0.217104   0.231380   0.085352
 O039   O   0.109620  -0.100193   0.091905
 O040   O  -0.111655   0.117464  -0.090507
 O041   O  -0.368966   0.381144  -0.416868
 O042   O   0.412314  -0.411186   0.401973
 O043   O   0.432129   0.232434   0.100436
 O044   O   0.223437   0.453349  -0.089544
 O045   O  -0.445694  -0.215612  -0.088437
 O046   O  -0.221653  -0.423391   0.092507
 O047   O  -0.039562  -0.282744  -0.414431
 O048   O  -0.252702  -0.068497   0.407807
 O049   O   0.075797   0.259961   0.403233
 O050   O   0.296615   0.053515  -0.413830
 O051   O   0.245157   0.095661  -0.068125
 O052   O   0.097557   0.247308   0.067972
 O053   O  -0.236879  -0.086502   0.069031
 O054   O  -0.086752  -0.235159  -0.071459
 O055   O  -0.247611  -0.412093   0.428513
 O056   O  -0.383263  -0.273821  -0.413020
 O057   O   0.286868   0.396423  -0.414143
 O058   O   0.418522   0.256308   0.426654
 O059   O  -0.435631   0.441694  -0.077458
 O060   O   0.423665  -0.416667   0.070720
 O061   O   0.082257  -0.069251   0.435411
 O062   O  -0.066390   0.080611  -0.422303
Na063  Na   0.325982   0.321448   0.283188
\end{verbatim}
\normalsize

KMO structure with two Na$^+$ intercalants (Na$_{2i}$)
\footnotesize
\begin{verbatim}
_cell_length_a                         8.882523
_cell_length_b                         8.893349
_cell_length_c                         12.573381
_cell_angle_alpha                      101.005522
_cell_angle_beta                       79.401083
_cell_angle_gamma                      123.105238
_symmetry_space_group_name_H-M         'P 1'
_symmetry_Int_Tables_number            1

loop_
_symmetry_equiv_pos_as_xyz
   'x, y, z'

loop_
   _atom_site_label
   _atom_site_type_symbol
   _atom_site_fract_x
   _atom_site_fract_y
   _atom_site_fract_z
 K001   K  -0.013048   0.357312   0.246636
 K002   K  -0.453617  -0.207325   0.240659
 K003   K   0.161469  -0.173840  -0.244917
 K004   K  -0.370861  -0.024709  -0.252944
 K005   K   0.035974   0.365175  -0.256882
 K006   K  -0.476563   0.142541   0.249057
 K007   K  -0.289967  -0.349237  -0.249540
 K008   K  -0.040537  -0.265651   0.259735
MN009  MN  -0.485596  -0.175509   0.488471
MN010  MN  -0.151366   0.496543  -0.495963
MN011  MN  -0.487526   0.159310  -0.494410
MN012  MN   0.182893   0.494689   0.486986
MN013  MN  -0.012766   0.332014  -0.006484
MN014  MN   0.322589   0.003014   0.012952
MN015  MN  -0.009041  -0.334034   0.006808
MN016  MN  -0.342078   0.000007  -0.000875
MN017  MN  -0.153685   0.161281   0.487054
MN018  MN   0.180825  -0.170618   0.499491
MN019  MN   0.324265  -0.331484  -0.004109
MN020  MN  -0.346042   0.335952   0.011809
MN021  MN  -0.005162   0.000040  -0.010890
MN022  MN  -0.486708   0.495327  -0.489084
MN023  MN   0.322446   0.333085  -0.009443
MN024  MN  -0.343530  -0.338360  -0.004702
MN025  MN  -0.156695  -0.168954  -0.490521
MN026  MN   0.179328   0.166277  -0.488666
 O027   O   0.400728  -0.058490   0.412814
 O028   O  -0.052485   0.396572  -0.431730
 O029   O  -0.385033   0.064538  -0.430335
 O030   O   0.064943  -0.390672   0.411580
 O031   O  -0.111504   0.431724  -0.072849
 O032   O   0.437933  -0.112550   0.088158
 O033   O   0.089433  -0.432592   0.080818
 O034   O  -0.444212   0.099173  -0.065706
 O035   O  -0.256902   0.260717   0.416955
 O036   O   0.280506  -0.269434  -0.432762
 O037   O   0.225588  -0.235346  -0.072756
 O038   O  -0.227786   0.222854   0.086980
 O039   O   0.091403  -0.106889   0.087443
 O040   O  -0.123490   0.111290  -0.094350
 O041   O  -0.374733   0.381090  -0.406302
 O042   O   0.410566  -0.410639   0.404251
 O043   O   0.423313   0.236950   0.097620
 O044   O   0.211235   0.446050  -0.091490
 O045   O  -0.454599  -0.223192  -0.090179
 O046   O  -0.233161  -0.430964   0.090570
 O047   O  -0.040428  -0.281937  -0.409501
 O048   O  -0.254694  -0.065344   0.407758
 O049   O   0.073793   0.262178   0.404653
 O050   O   0.294088   0.052470  -0.406096
 O051   O   0.223040   0.101968  -0.093383
 O052   O   0.086464   0.238699   0.065736
 O053   O  -0.240884  -0.096676   0.066087
 O054   O  -0.098455  -0.247579  -0.079872
 O055   O  -0.251191  -0.412433   0.433066
 O056   O  -0.382933  -0.269685  -0.410949
 O057   O   0.283902   0.392909  -0.408581
 O058   O   0.416795   0.257659   0.431818
 O059   O  -0.447628   0.435455  -0.090212
 O060   O   0.415162  -0.430284   0.070737
 O061   O   0.081767  -0.069776   0.435841
 O062   O  -0.068146   0.078827  -0.421847
Na063  Na   0.325041   0.333652   0.269947
Na064  Na   0.380620   0.264303  -0.250064
\end{verbatim}
\normalsize

KMO structure with three Na$^+$ intercalants (Na$_{3i}$)
\footnotesize
\begin{verbatim}
_cell_length_a                         9.000514
_cell_length_b                         8.935277
_cell_length_c                         12.432702
_cell_angle_alpha                      101.584522
_cell_angle_beta                       79.088014
_cell_angle_gamma                      123.717746
_symmetry_space_group_name_H-M         'P 1'
_symmetry_Int_Tables_number            1

loop_
_symmetry_equiv_pos_as_xyz
   'x, y, z'

loop_
   _atom_site_label
   _atom_site_type_symbol
   _atom_site_fract_x
   _atom_site_fract_y
   _atom_site_fract_z
 K001   K  -0.017148   0.342376   0.252187
 K002   K  -0.451748  -0.211393   0.241709
 K003   K   0.150440  -0.170098  -0.246643
 K004   K   0.495092  -0.125038  -0.254832
 K005   K   0.028653   0.332292  -0.260440
 K006   K  -0.476319   0.141816   0.251788
 K007   K  -0.189408  -0.170824  -0.247566
 K008   K  -0.033279  -0.276535   0.259671
MN009  MN  -0.479818  -0.170399   0.490280
MN010  MN  -0.149457  -0.499737  -0.495310
MN011  MN  -0.482238   0.165199  -0.490692
MN012  MN   0.185915   0.497669   0.492025
MN013  MN  -0.009461   0.328779  -0.008897
MN014  MN   0.327120   0.000227   0.009915
MN015  MN  -0.006087  -0.337569   0.003118
MN016  MN  -0.339053  -0.003056  -0.004527
MN017  MN  -0.148464   0.163006   0.491332
MN018  MN   0.186129  -0.167034  -0.498986
MN019  MN   0.327057  -0.335299  -0.007296
MN020  MN  -0.342032   0.332181   0.010258
MN021  MN  -0.002018  -0.002478  -0.007148
MN022  MN  -0.481741  -0.499155  -0.486381
MN023  MN   0.326162   0.329145  -0.013303
MN024  MN  -0.340769  -0.342913  -0.012948
MN025  MN  -0.149195  -0.165350  -0.493034
MN026  MN   0.184001   0.165849  -0.484486
 O027   O   0.407774  -0.054828   0.414886
 O028   O  -0.052176   0.403160  -0.424482
 O029   O  -0.382167   0.071874  -0.422294
 O030   O   0.068139  -0.391409   0.412293
 O031   O  -0.110092   0.426567  -0.079153
 O032   O   0.441618  -0.114731   0.084929
 O033   O   0.092426  -0.435276   0.077156
 O034   O  -0.438907   0.095554  -0.071767
 O035   O  -0.265996   0.274474   0.416360
 O036   O   0.285870  -0.264470  -0.428550
 O037   O   0.229755  -0.238438  -0.077647
 O038   O  -0.224375   0.220966   0.084458
 O039   O   0.095913  -0.110214   0.085321
 O040   O  -0.120246   0.106903  -0.096732
 O041   O  -0.370500   0.391034  -0.398976
 O042   O   0.412439  -0.408769   0.407050
 O043   O   0.427999   0.235209   0.094197
 O044   O   0.213196   0.440631  -0.096226
 O045   O  -0.453170  -0.228991  -0.094268
 O046   O  -0.227999  -0.433007   0.086815
 O047   O  -0.033592  -0.277311  -0.407798
 O048   O  -0.252480  -0.068324   0.409358
 O049   O   0.084593   0.265969   0.409670
 O050   O   0.298505   0.055267  -0.401218
 O051   O   0.221629   0.096902  -0.092777
 O052   O   0.089985   0.240067   0.064292
 O053   O  -0.240729  -0.101127   0.062771
 O054   O  -0.096288  -0.251653  -0.082838
 O055   O  -0.247721  -0.408898   0.433037
 O056   O  -0.377319  -0.265051  -0.409584
 O057   O   0.290909   0.399573  -0.401379
 O058   O   0.418921   0.261270   0.435261
 O059   O  -0.443505   0.428891  -0.099168
 O060   O   0.420568  -0.432911   0.064577
 O061   O   0.091773  -0.069152   0.433409
 O062   O  -0.048009   0.062756  -0.405956
Na063  Na   0.329464   0.330745   0.281901
Na064  Na   0.374758   0.326770  -0.260808
Na065  Na  -0.280809   0.361564  -0.252345
\end{verbatim}
\normalsize

KMO structure with four Na$^+$ intercalants (Na$_{4i}$)
\footnotesize
\begin{verbatim}
_cell_length_a                         9.035862
_cell_length_b                         9.035562
_cell_length_c                         12.426189
_cell_angle_alpha                      102.154128
_cell_angle_beta                       77.971226
_cell_angle_gamma                      124.078582
_symmetry_space_group_name_H-M         'P 1'
_symmetry_Int_Tables_number            1

loop_
_symmetry_equiv_pos_as_xyz
   'x, y, z'

loop_
   _atom_site_label
   _atom_site_type_symbol
   _atom_site_fract_x
   _atom_site_fract_y
   _atom_site_fract_z
 K001   K  -0.025951   0.326898   0.248255
 K002   K  -0.451598  -0.227092   0.237124
 K003   K   0.187575  -0.204031  -0.242268
 K004   K  -0.360171  -0.033275  -0.255505
 K005   K   0.015215   0.342871  -0.258635
 K006   K  -0.490116   0.128331   0.250634
 K007   K   0.008199  -0.020494  -0.253223
 K008   K  -0.047388  -0.287402   0.260046
MN009  MN  -0.495218  -0.193713   0.490128
MN010  MN  -0.164789   0.478021  -0.494642
MN011  MN  -0.498419   0.142899  -0.492598
MN012  MN   0.171691   0.475239   0.486048
MN013  MN  -0.013442   0.327602  -0.006123
MN014  MN   0.318092  -0.004459   0.015740
MN015  MN  -0.016217  -0.338314   0.009251
MN016  MN  -0.345780  -0.006221  -0.004136
MN017  MN  -0.163550   0.141183   0.486682
MN018  MN   0.170231  -0.189914  -0.496721
MN019  MN   0.319731  -0.341157  -0.001777
MN020  MN  -0.348575   0.330890   0.012601
MN021  MN  -0.007762  -0.009988  -0.011093
MN022  MN  -0.498339   0.478147  -0.485770
MN023  MN   0.322225   0.324607  -0.020409
MN024  MN  -0.343594  -0.341385  -0.021794
MN025  MN  -0.168165  -0.187839  -0.483947
MN026  MN   0.167568   0.147950  -0.485907
 O027   O   0.391676  -0.075444   0.414166
 O028   O  -0.067048   0.380372  -0.430445
 O029   O  -0.402234   0.047533  -0.426610
 O030   O   0.053248  -0.409425   0.412516
 O031   O  -0.113388   0.425628  -0.072309
 O032   O   0.436201  -0.121286   0.089244
 O033   O   0.097470  -0.455180   0.087842
 O034   O  -0.448154   0.092301  -0.068157
 O035   O  -0.280312   0.255764   0.415070
 O036   O   0.268765  -0.290630  -0.433502
 O037   O   0.225340  -0.243041  -0.065379
 O038   O  -0.232775   0.215979   0.086720
 O039   O   0.083878  -0.104003   0.088457
 O040   O  -0.129395   0.108707  -0.098733
 O041   O  -0.386711   0.366355  -0.396907
 O042   O   0.403982  -0.429263   0.406167
 O043   O   0.418515   0.232237   0.094211
 O044   O   0.208818   0.439515  -0.098597
 O045   O  -0.459616  -0.229501  -0.097585
 O046   O  -0.250630  -0.435200   0.086078
 O047   O  -0.049651  -0.302428  -0.399179
 O048   O  -0.264569  -0.092843   0.411632
 O049   O   0.073599   0.244048   0.405466
 O050   O   0.282861   0.029983  -0.398692
 O051   O   0.224067   0.090090  -0.100037
 O052   O   0.086100   0.229965   0.058987
 O053   O  -0.246663  -0.102415   0.059289
 O054   O  -0.109308  -0.241341  -0.102568
 O055   O  -0.260183  -0.426101   0.438713
 O056   O  -0.399275  -0.292583  -0.400358
 O057   O   0.272754   0.378390  -0.401590
 O058   O   0.403973   0.239527   0.437302
 O059   O  -0.448846   0.428121  -0.111510
 O060   O   0.419130  -0.440642   0.059880
 O061   O   0.072881  -0.091498   0.441085
 O062   O  -0.068947   0.050761  -0.407292
Na063  Na   0.317317   0.319567   0.268349
Na064  Na   0.440849  -0.462368  -0.242699
Na065  Na  -0.214920  -0.467286  -0.249223
Na066  Na   0.451002   0.194769  -0.248096
\end{verbatim}
\normalsize

KMO structure with five Na$^+$ intercalants (Na$_{5i}$)
\footnotesize
\begin{verbatim}
_cell_length_a                         9.097981
_cell_length_b                         9.033969
_cell_length_c                         12.206746
_cell_angle_alpha                      101.604905
_cell_angle_beta                       79.907502
_cell_angle_gamma                      124.032754
_symmetry_space_group_name_H-M         'P 1'
_symmetry_Int_Tables_number            1

loop_
_symmetry_equiv_pos_as_xyz
   'x, y, z'

loop_
   _atom_site_label
   _atom_site_type_symbol
   _atom_site_fract_x
   _atom_site_fract_y
   _atom_site_fract_z
 K001   K  -0.027076   0.322343   0.249619
 K002   K  -0.471078  -0.208040   0.241373
 K003   K   0.146358  -0.203417  -0.240332
 K004   K  -0.472603  -0.080653  -0.248025
 K005   K   0.005215   0.297459  -0.259082
 K006   K  -0.376467   0.312745   0.252976
 K007   K  -0.157029  -0.137774  -0.243758
 K008   K  -0.112838  -0.166233   0.249176
MN009  MN   0.488165  -0.213481   0.490987
MN010  MN  -0.177101   0.458995  -0.497299
MN011  MN   0.489301   0.123027  -0.499605
MN012  MN   0.158482   0.456511   0.488424
MN013  MN   0.018301   0.340703  -0.006249
MN014  MN   0.352865   0.013768   0.014324
MN015  MN   0.019084  -0.323877   0.012177
MN016  MN  -0.318675   0.014985   0.006374
MN017  MN  -0.176998   0.123317   0.488782
MN018  MN   0.155780  -0.210231   0.499729
MN019  MN   0.348790  -0.324724   0.000776
MN020  MN  -0.316098   0.347097   0.005471
MN021  MN   0.013269   0.004968  -0.007857
MN022  MN   0.491223   0.459324  -0.487850
MN023  MN   0.344822   0.335662  -0.019151
MN024  MN  -0.322449  -0.331240  -0.012174
MN025  MN  -0.177166  -0.208016  -0.491060
MN026  MN   0.158947   0.125111  -0.491720
 O027   O   0.368909  -0.100884   0.407541
 O028   O  -0.076523   0.364307  -0.428041
 O029   O  -0.410855   0.030523  -0.431543
 O030   O   0.037596  -0.431652   0.410837
 O031   O  -0.082673   0.440968  -0.077912
 O032   O   0.478679  -0.066270   0.105057
 O033   O   0.131417  -0.437431   0.091328
 O034   O  -0.415225   0.109717  -0.069222
 O035   O  -0.292904   0.234408   0.411614
 O036   O   0.258334  -0.305495  -0.429622
 O037   O   0.251673  -0.223235  -0.070189
 O038   O  -0.196644   0.242781   0.087099
 O039   O   0.128527  -0.088365   0.099058
 O040   O  -0.106413   0.113445  -0.090804
 O041   O  -0.395744   0.346708  -0.402683
 O042   O   0.383023  -0.448673   0.405278
 O043   O   0.453066   0.253443   0.086963
 O044   O   0.219349   0.422157  -0.106158
 O045   O  -0.431548  -0.210691  -0.090938
 O046   O  -0.212652  -0.425083   0.089668
 O047   O  -0.060680  -0.319381  -0.404417
 O048   O  -0.286027  -0.112619   0.413016
 O049   O   0.054800   0.222474   0.405314
 O050   O   0.272878   0.012398  -0.406545
 O051   O   0.234160   0.096838  -0.097869
 O052   O   0.116024   0.244895   0.069111
 O053   O  -0.219115  -0.086804   0.074744
 O054   O  -0.093746  -0.232763  -0.088740
 O055   O  -0.276461  -0.446720   0.432142
 O056   O  -0.403434  -0.306653  -0.401340
 O057   O   0.270384   0.357609  -0.401427
 O058   O   0.391120   0.219509   0.429363
 O059   O  -0.432358   0.440717  -0.112740
 O060   O   0.448699  -0.431785   0.066607
 O061   O   0.057894  -0.116971   0.427574
 O062   O  -0.069094   0.027436  -0.412608
Na063  Na   0.289229   0.305296   0.240167
Na064  Na   0.440536  -0.444029  -0.236268
Na065  Na  -0.230451   0.445430  -0.246914
Na066  Na   0.357367   0.222863  -0.258340
Na067  Na   0.225602  -0.166519   0.238822
\end{verbatim}
\normalsize

KMO structure with six Na$^+$ intercalants (Na$_{6i}$)
\footnotesize
\begin{verbatim}
_cell_length_a                         9.151285
_cell_length_b                         9.164955
_cell_length_c                         12.107997
_cell_angle_alpha                      100.997862
_cell_angle_beta                       79.300379
_cell_angle_gamma                      124.637943
_symmetry_space_group_name_H-M         'P 1'
_symmetry_Int_Tables_number            1

loop_
_symmetry_equiv_pos_as_xyz
   'x, y, z'

loop_
   _atom_site_label
   _atom_site_type_symbol
   _atom_site_fract_x
   _atom_site_fract_y
   _atom_site_fract_z
 K001   K  -0.002172   0.336118   0.255524
 K002   K  -0.457987  -0.213472   0.242524
 K003   K   0.130584  -0.208079  -0.249099
 K004   K  -0.492929  -0.101517  -0.250168
 K005   K   0.012567   0.294277  -0.254963
 K006   K  -0.464916   0.150888   0.247649
 K007   K  -0.166312  -0.147470  -0.250027
 K008   K  -0.019569  -0.006002   0.252563
MN009  MN   0.496409  -0.184740   0.490516
MN010  MN  -0.170165   0.482626   0.498998
MN011  MN   0.498191   0.148805  -0.497052
MN012  MN   0.167173   0.480103   0.496292
MN013  MN   0.000739   0.318025   0.000602
MN014  MN   0.330002  -0.018090   0.008263
MN015  MN   0.001314  -0.349529   0.007503
MN016  MN  -0.332711  -0.015560   0.001701
MN017  MN  -0.166590   0.151530   0.498899
MN018  MN   0.162787  -0.187147   0.491178
MN019  MN   0.332343  -0.350524   0.001440
MN020  MN  -0.334057   0.317219   0.010218
MN021  MN  -0.003173  -0.017773  -0.013134
MN022  MN   0.497340   0.482885  -0.492511
MN023  MN   0.333216   0.312561  -0.009972
MN024  MN  -0.336222  -0.358597  -0.007763
MN025  MN  -0.173785  -0.187232   0.497968
MN026  MN   0.167393   0.145342  -0.491758
 O027   O   0.386841  -0.070959   0.402812
 O028   O  -0.059759   0.381516  -0.415081
 O029   O  -0.408201   0.053631  -0.425351
 O030   O   0.053687  -0.398821   0.395305
 O031   O  -0.100070   0.412881  -0.072960
 O032   O   0.447554  -0.125189   0.095144
 O033   O   0.107316  -0.462683   0.096168
 O034   O  -0.428406   0.083600  -0.067736
 O035   O  -0.281418   0.259640   0.411515
 O036   O   0.262797  -0.280727  -0.433175
 O037   O   0.228249  -0.242125  -0.080514
 O038   O  -0.216114   0.206359   0.093413
 O039   O   0.108588  -0.128848   0.098860
 O040   O  -0.115309   0.097174  -0.090448
 O041   O  -0.398575   0.364739  -0.400018
 O042   O   0.390406  -0.418304   0.406965
 O043   O   0.433360   0.214878   0.091438
 O044   O   0.219867   0.425720  -0.092347
 O045   O  -0.442386  -0.235635  -0.093049
 O046   O  -0.232056  -0.455458   0.094260
 O047   O  -0.054552  -0.294322  -0.415222
 O048   O  -0.272912  -0.075741   0.411755
 O049   O   0.066496   0.250585   0.415000
 O050   O   0.273708   0.028560  -0.409953
 O051   O   0.223532   0.081882  -0.096773
 O052   O   0.095724   0.216115   0.070281
 O053   O  -0.233811  -0.112651   0.071051
 O054   O  -0.109032  -0.253585  -0.090511
 O055   O  -0.285348  -0.413473   0.412876
 O056   O  -0.399978  -0.288524  -0.404842
 O057   O   0.277107   0.375327  -0.400277
 O058   O   0.401093   0.244843   0.427984
 O059   O  -0.442584   0.415479  -0.105192
 O060   O   0.441745  -0.466219   0.081591
 O061   O   0.070617  -0.091116   0.416993
 O062   O  -0.063355   0.037874  -0.413468
Na063  Na   0.426205   0.410927   0.245961
Na064  Na   0.439720  -0.457753  -0.242358
Na065  Na  -0.237764   0.422404  -0.250891
Na066  Na   0.371286   0.214764  -0.256101
Na067  Na   0.212424  -0.204720   0.240288
Na068  Na  -0.136147  -0.433064   0.256040
\end{verbatim}
\normalsize

KMO structure with seven Na$^+$ intercalants (Na$_{7i}$)
\footnotesize
\begin{verbatim}
_cell_length_a                         9.251462
_cell_length_b                         9.266501
_cell_length_c                         11.661336
_cell_angle_alpha                      93.193040
_cell_angle_beta                       85.590487
_cell_angle_gamma                      125.000061
_symmetry_space_group_name_H-M         'P 1'
_symmetry_Int_Tables_number            1

loop_
_symmetry_equiv_pos_as_xyz
   'x, y, z'

loop_
   _atom_site_label
   _atom_site_type_symbol
   _atom_site_fract_x
   _atom_site_fract_y
   _atom_site_fract_z
 K001   K  -0.088843   0.387375   0.251193
 K002   K  -0.410880  -0.304680   0.248009
 K003   K   0.172955  -0.178267  -0.247690
 K004   K  -0.405204  -0.048660  -0.257168
 K005   K   0.003774   0.282890  -0.254528
 K006   K  -0.420082   0.058638   0.249501
 K007   K  -0.163572  -0.168803  -0.248620
 K008   K  -0.074605   0.045230   0.249673
MN009  MN   0.457097  -0.178029   0.489963
MN010  MN  -0.204281   0.493124   0.496417
MN011  MN   0.460191   0.159688   0.497060
MN012  MN   0.130483   0.491156   0.495065
MN013  MN   0.040567   0.281428   0.004282
MN014  MN   0.366997  -0.053456   0.013879
MN015  MN   0.038486  -0.381453   0.012875
MN016  MN  -0.298354  -0.054383   0.003343
MN017  MN  -0.205651   0.162813   0.493416
MN018  MN   0.125934  -0.174527   0.494912
MN019  MN   0.370791  -0.387896   0.001560
MN020  MN  -0.296456   0.283307   0.013034
MN021  MN   0.035665  -0.050242  -0.007173
MN022  MN   0.469637   0.490219  -0.496917
MN023  MN   0.374222   0.279549  -0.005965
MN024  MN  -0.298283  -0.393359  -0.006556
MN025  MN  -0.205711  -0.172128   0.497003
MN026  MN   0.130053   0.155507  -0.497430
 O027   O   0.340379  -0.049881   0.400183
 O028   O  -0.093947   0.371723  -0.415892
 O029   O  -0.444236   0.054324  -0.431331
 O030   O   0.018545  -0.380127   0.399682
 O031   O  -0.063706   0.388855  -0.063556
 O032   O   0.490441  -0.179990   0.099310
 O033   O   0.161854   0.482820   0.102830
 O034   O  -0.396422   0.057522  -0.061133
 O035   O  -0.327134   0.282917   0.406986
 O036   O   0.254576  -0.302435  -0.412402
 O037   O   0.257286  -0.265101  -0.074196
 O038   O  -0.172652   0.152180   0.096860
 O039   O   0.148961  -0.169742   0.112214
 O040   O  -0.084480   0.072148  -0.088737
 O041   O  -0.427533   0.363755  -0.407173
 O042   O   0.341683  -0.392136   0.395304
 O043   O   0.476919   0.164400   0.096555
 O044   O   0.246827   0.400257  -0.095296
 O045   O  -0.416546  -0.255931  -0.094779
 O046   O  -0.184061  -0.497921   0.097696
 O047   O  -0.084765  -0.304238  -0.408396
 O048   O  -0.320894  -0.052908   0.404397
 O049   O   0.021064   0.280397   0.403575
 O050   O   0.245362   0.038407  -0.411154
 O051   O   0.256066   0.063052  -0.090952
 O052   O   0.142573   0.174874   0.078101
 O053   O  -0.194975  -0.162931   0.072832
 O054   O  -0.071529  -0.272804  -0.086673
 O055   O  -0.313475  -0.386475   0.411029
 O056   O  -0.423332  -0.291319  -0.409858
 O057   O   0.252667   0.369170  -0.405914
 O058   O   0.357917   0.265958   0.415036
 O059   O  -0.403651   0.392802  -0.096861
 O060   O   0.483202   0.488307   0.081103
 O061   O   0.016664  -0.055719   0.405486
 O062   O  -0.095989   0.041717  -0.413541
Na063  Na   0.262184   0.385482   0.250090
Na064  Na   0.367270  -0.419250  -0.260963
Na065  Na  -0.231516   0.444934  -0.262872
Na066  Na   0.352295   0.242695  -0.261544
Na067  Na   0.275041   0.070416   0.259035
Na068  Na  -0.093624  -0.304462   0.262172
Na069  Na   0.250220  -0.282466   0.255664
\end{verbatim}
\normalsize

KMO structure with eight Na$^+$ intercalants (Na$_{8i}$)
\footnotesize
\begin{verbatim}
_cell_length_a                         9.373304
_cell_length_b                         9.360209
_cell_length_c                         11.625006
_cell_angle_alpha                      93.601128
_cell_angle_beta                       79.662622
_cell_angle_gamma                      125.379741
_symmetry_space_group_name_H-M         'P 1'
_symmetry_Int_Tables_number            1

loop_
_symmetry_equiv_pos_as_xyz
   'x, y, z'

loop_
   _atom_site_label
   _atom_site_type_symbol
   _atom_site_fract_x
   _atom_site_fract_y
   _atom_site_fract_z
 K001   K  -0.090707   0.387588   0.251997
 K002   K  -0.417772  -0.307027   0.250768
 K003   K   0.247224  -0.118733  -0.252226
 K004   K  -0.424291  -0.127249  -0.250359
 K005   K  -0.083122   0.225992  -0.251154
 K006   K  -0.420453   0.058220   0.252284
 K007   K  -0.099714  -0.129399  -0.252863
 K008   K  -0.076972   0.049231   0.251398
MN009  MN   0.414914  -0.203508  -0.498705
MN010  MN  -0.252285   0.467194  -0.498655
MN011  MN   0.415478   0.135252  -0.499965
MN012  MN   0.081265   0.467351  -0.498828
MN013  MN   0.084680   0.303810  -0.001261
MN014  MN   0.408920  -0.036955   0.010148
MN015  MN   0.078542  -0.362940   0.008748
MN016  MN  -0.254641  -0.034818   0.002195
MN017  MN  -0.252446   0.134326  -0.498563
MN018  MN   0.081157  -0.195889   0.496242
MN019  MN   0.413255  -0.369194   0.000641
MN020  MN  -0.254559   0.300640   0.007225
MN021  MN   0.080836  -0.025864  -0.009428
MN022  MN   0.420906   0.463126   0.498916
MN023  MN   0.421340   0.303797  -0.011574
MN024  MN  -0.252888  -0.371382  -0.007825
MN025  MN  -0.249208  -0.203171   0.498996
MN026  MN   0.086134   0.134844   0.496283
 O027   O   0.313007  -0.069334   0.403820
 O028   O  -0.156563   0.339670  -0.402470
 O029   O  -0.484403   0.005642  -0.405599
 O030   O  -0.011303  -0.397227   0.404013
 O031   O  -0.005062   0.418047  -0.072737
 O032   O  -0.485003  -0.166102   0.097284
 O033   O   0.185732   0.494227   0.099518
 O034   O  -0.342363   0.079345  -0.072019
 O035   O  -0.350157   0.263362   0.413800
 O036   O   0.182256  -0.342683  -0.397748
 O037   O   0.309869  -0.239017  -0.081042
 O038   O  -0.147331   0.166527   0.092340
 O039   O   0.173389  -0.158905   0.106117
 O040   O  -0.019899   0.102278  -0.093811
 O041   O  -0.490757   0.333733  -0.404012
 O042   O   0.309213  -0.400630   0.400507
 O043   O  -0.495085   0.173414   0.090377
 O044   O   0.304093   0.432221  -0.101245
 O045   O  -0.353939  -0.231929  -0.087822
 O046   O  -0.158341  -0.489184   0.093057
 O047   O  -0.148165  -0.332226  -0.407477
 O048   O  -0.352795  -0.070797   0.408210
 O049   O  -0.014470   0.267228   0.407840
 O050   O   0.181237   0.008582  -0.408284
 O051   O   0.317140   0.100101  -0.099780
 O052   O   0.173596   0.191472   0.073077
 O053   O  -0.162690  -0.145420   0.073862
 O054   O  -0.015235  -0.240970  -0.090637
 O055   O  -0.341400  -0.405144   0.408592
 O056   O  -0.487604  -0.332981  -0.411666
 O057   O   0.188629   0.335999  -0.405075
 O058   O   0.314471   0.267593   0.397965
 O059   O  -0.336932   0.421884  -0.101377
 O060   O  -0.487374  -0.499269   0.078161
 O061   O  -0.010751  -0.073709   0.401222
 O062   O  -0.149652   0.004783  -0.413140
Na063  Na   0.254882   0.384759   0.257120
Na064  Na   0.266310  -0.454888  -0.263077
Na065  Na  -0.096806  -0.447105  -0.265433
Na066  Na   0.276180   0.247339  -0.259857
Na067  Na   0.274788   0.071771   0.260256
Na068  Na  -0.096125  -0.304779   0.263809
Na069  Na   0.246543  -0.281134   0.263151
Na070  Na  -0.419344   0.240937  -0.265040
\end{verbatim}
\normalsize

KMO structure with nine Na$^+$ intercalants (Na$_{9i}$)
\footnotesize
\begin{verbatim}
_cell_length_a                         9.465042
_cell_length_b                         9.460521
_cell_length_c                         11.460976
_cell_angle_alpha                      93.192990
_cell_angle_beta                       80.146627
_cell_angle_gamma                      125.954183
_symmetry_space_group_name_H-M         'P 1'
_symmetry_Int_Tables_number            1

loop_
_symmetry_equiv_pos_as_xyz
   'x, y, z'

loop_
   _atom_site_label
   _atom_site_type_symbol
   _atom_site_fract_x
   _atom_site_fract_y
   _atom_site_fract_z
 K001   K  -0.081566   0.385655   0.252828
 K002   K  -0.420677  -0.287026   0.249608
 K003   K   0.249461  -0.114492  -0.250619
 K004   K  -0.423534  -0.131182  -0.248627
 K005   K  -0.081225   0.222519  -0.251397
 K006   K  -0.420300   0.050766   0.250844
 K007   K  -0.098015  -0.131481  -0.250404
 K008   K  -0.077863   0.054040   0.251321
MN009  MN   0.414036  -0.203657  -0.497734
MN010  MN  -0.252039   0.467013   0.499628
MN011  MN   0.415526   0.133405   0.499011
MN012  MN   0.081062   0.468038  -0.498052
MN013  MN   0.084777   0.303222  -0.001191
MN014  MN   0.409656  -0.032507   0.007582
MN015  MN   0.080161  -0.365468   0.008550
MN016  MN  -0.253150  -0.033957   0.001621
MN017  MN  -0.250954   0.133125  -0.496433
MN018  MN   0.080453  -0.196911   0.495350
MN019  MN   0.414326  -0.366690   0.000659
MN020  MN  -0.257754   0.301231   0.002819
MN021  MN   0.081117  -0.029417  -0.004520
MN022  MN   0.420641   0.461840   0.496721
MN023  MN   0.419024   0.306859  -0.006905
MN024  MN  -0.249875  -0.369623  -0.000749
MN025  MN  -0.249412  -0.205286   0.496778
MN026  MN   0.087014   0.133131   0.494845
 O027   O   0.311475  -0.069766   0.402123
 O028   O  -0.157179   0.339341  -0.403647
 O029   O  -0.483801   0.006392  -0.406702
 O030   O  -0.013883  -0.395994   0.401889
 O031   O  -0.002358   0.419660  -0.076550
 O032   O  -0.491062  -0.163422   0.093603
 O033   O   0.186990   0.495311   0.100073
 O034   O  -0.359476   0.102832  -0.090412
 O035   O  -0.355134   0.269764   0.403344
 O036   O   0.183027  -0.342863  -0.396215
 O037   O   0.308736  -0.239172  -0.083538
 O038   O  -0.148189   0.170398   0.089982
 O039   O   0.178298  -0.163047   0.107650
 O040   O  -0.014190   0.102216  -0.092836
 O041   O  -0.488069   0.331631  -0.405460
 O042   O   0.306588  -0.399544   0.400369
 O043   O  -0.497173   0.174333   0.097696
 O044   O   0.307709   0.434782  -0.099379
 O045   O  -0.351275  -0.237950  -0.084149
 O046   O  -0.157999  -0.498386   0.098368
 O047   O  -0.148989  -0.333532  -0.408637
 O048   O  -0.352419  -0.069028   0.410257
 O049   O  -0.014331   0.266503   0.410545
 O050   O   0.181972   0.008415  -0.410453
 O051   O   0.314413   0.106312  -0.098452
 O052   O   0.173415   0.192793   0.078316
 O053   O  -0.144575  -0.164682   0.092768
 O054   O  -0.014215  -0.243839  -0.089236
 O055   O  -0.344435  -0.406463   0.400236
 O056   O  -0.486371  -0.335643  -0.413978
 O057   O   0.191505   0.333409  -0.406758
 O058   O   0.315952   0.268441   0.390442
 O059   O  -0.340993   0.428967  -0.096590
 O060   O  -0.486265   0.495302   0.093581
 O061   O  -0.012348  -0.074017   0.400156
 O062   O  -0.148816   0.001745  -0.414273
Na063  Na   0.252072   0.387927   0.252241
Na064  Na   0.262139  -0.456167  -0.258244
Na065  Na  -0.100080  -0.448335  -0.259858
Na066  Na   0.283673   0.252000  -0.255871
Na067  Na   0.250453   0.047284   0.256884
Na068  Na  -0.083273  -0.285518   0.252743
Na069  Na   0.244821  -0.281444   0.256871
Na070  Na  -0.418979   0.233718  -0.254030
Na071  Na  -0.421287   0.383447   0.250131
\end{verbatim}
\normalsize

Fully-intercalated KMO structure with ten Na$^+$ intercalants (Na$_{10i}$)
\footnotesize
\begin{verbatim}
_cell_length_a                         9.565984
_cell_length_b                         9.569428
_cell_length_c                         11.336024
_cell_angle_alpha                      93.231024
_cell_angle_beta                       80.130750
_cell_angle_gamma                      126.532075
_symmetry_space_group_name_H-M         'P 1'
_symmetry_Int_Tables_number            1

loop_
_symmetry_equiv_pos_as_xyz
   'x, y, z'

loop_
   _atom_site_label
   _atom_site_type_symbol
   _atom_site_fract_x
   _atom_site_fract_y
   _atom_site_fract_z
 K001   K  -0.080804   0.388269   0.249638
 K002   K  -0.423048  -0.289237   0.249803
 K003   K   0.247614  -0.116210  -0.250025
 K004   K  -0.423782  -0.120901  -0.250464
 K005   K  -0.080352   0.221500  -0.249531
 K006   K  -0.420288   0.050354   0.249827
 K007   K  -0.082717  -0.117174  -0.250182
 K008   K  -0.081139   0.054972   0.249199
MN009  MN   0.410454  -0.201580  -0.499098
MN010  MN  -0.252406   0.466626  -0.497263
MN011  MN   0.414847   0.134932   0.499730
MN012  MN   0.079946   0.467998  -0.498169
MN013  MN   0.087564   0.302905  -0.003154
MN014  MN   0.412828  -0.032706   0.005272
MN015  MN   0.077791  -0.362624   0.003885
MN016  MN  -0.249009  -0.034193  -0.001325
MN017  MN  -0.251356   0.133441  -0.496859
MN018  MN   0.078560  -0.196380   0.496611
MN019  MN   0.415201  -0.366799  -0.002716
MN020  MN  -0.254816   0.306038  -0.001827
MN021  MN   0.084826  -0.030681   0.000844
MN022  MN   0.419597   0.461157   0.499353
MN023  MN   0.418859   0.300951  -0.002673
MN024  MN  -0.250328  -0.368818  -0.000142
MN025  MN  -0.250164  -0.203858   0.497360
MN026  MN   0.086553   0.134354   0.494115
 O027   O   0.309629  -0.069617   0.402926
 O028   O  -0.157595   0.337565  -0.403616
 O029   O  -0.485778   0.008127  -0.408723
 O030   O  -0.016295  -0.395499   0.404251
 O031   O  -0.017315   0.443778  -0.095863
 O032   O  -0.485429  -0.162404   0.089857
 O033   O   0.190965   0.497127   0.098849
 O034   O  -0.360631   0.105991  -0.096679
 O035   O  -0.356249   0.271918   0.404446
 O036   O   0.182580  -0.342150  -0.396664
 O037   O   0.310347  -0.238846  -0.086677
 O038   O  -0.149976   0.172399   0.084321
 O039   O   0.182069  -0.166411   0.107479
 O040   O  -0.010378   0.098443  -0.092644
 O041   O  -0.488687   0.332086  -0.398196
 O042   O   0.306347  -0.398614   0.401696
 O043   O  -0.489029   0.175283   0.095426
 O044   O   0.315733   0.437268  -0.106111
 O045   O  -0.350626  -0.238734  -0.092732
 O046   O  -0.159541  -0.495799   0.097606
 O047   O  -0.151617  -0.337785  -0.400276
 O048   O  -0.354360  -0.071916   0.412253
 O049   O  -0.015017   0.266888   0.410045
 O050   O   0.182802   0.009707  -0.410588
 O051   O   0.314850   0.104841  -0.099381
 O052   O   0.189838   0.171573   0.096430
 O053   O  -0.146944  -0.166873   0.095256
 O054   O  -0.015594  -0.239020  -0.089484
 O055   O  -0.345481  -0.406374   0.403588
 O056   O  -0.482956  -0.341840  -0.403843
 O057   O   0.190863   0.331923  -0.405018
 O058   O   0.317299   0.269927   0.392681
 O059   O  -0.348296   0.436769  -0.104937
 O060   O  -0.484462   0.495288   0.094665
 O061   O  -0.014501  -0.074606   0.401159
 O062   O  -0.148533   0.002154  -0.413871
Na063  Na   0.253936   0.383799   0.247304
Na064  Na   0.249847  -0.453984  -0.251622
Na065  Na  -0.087796  -0.453435  -0.250051
Na066  Na   0.256336   0.223413  -0.251262
Na067  Na   0.249086   0.051464   0.248939
Na068  Na  -0.082626  -0.283760   0.250935
Na069  Na   0.245093  -0.282594   0.252160
Na070  Na  -0.418366   0.224291  -0.250209
Na071  Na  -0.420441  -0.456105  -0.250771
Na072  Na  -0.420594   0.383783   0.248399
\end{verbatim}
\normalsize

Transition state for \ce{K+} intercalation
\footnotesize
\begin{verbatim}
_cell_length_a                         8.611642
_cell_length_b                         8.616381
_cell_length_c                         11.626590
_cell_angle_alpha                      74.004417
_cell_angle_beta                       75.659798
_cell_angle_gamma                      59.985867
_cell_volume                           711.817634
_space_group_name_H-M_alt              'P 1'
_space_group_IT_number                 1

loop_
_space_group_symop_operation_xyz
   'x, y, z'

loop_
   _atom_site_label
   _atom_site_occupancy
   _atom_site_fract_x
   _atom_site_fract_y
   _atom_site_fract_z
   Mn1        1.0     0.846409     0.173239     0.204260    
   Mn2        1.0     0.822628     0.160321     0.791249    
   Mn3        1.0     0.844665     0.506344     0.208639    
   Mn4        1.0     0.822939     0.494199     0.789089    
   Mn5        1.0     0.844360     0.839275     0.209372    
   Mn6        1.0     0.822724     0.827077     0.789915    
   Mn7        1.0     0.177046     0.172668     0.211013    
   Mn8        1.0     0.155623     0.160361     0.791215    
   Mn9        1.0     0.177614     0.506310     0.208754    
   Mn10       1.0     0.153476     0.493552     0.795652    
   Mn11       1.0     0.177271     0.839341     0.210243    
   Mn12       1.0     0.155887     0.827239     0.790524    
   Mn13       1.0     0.511261     0.173247     0.204146    
   Mn14       1.0     0.487732     0.157272     0.799932    
   Mn15       1.0     0.511936     0.509605     0.199923    
   Mn16       1.0     0.488539     0.493509     0.795773    
   Mn17       1.0     0.510334     0.839562     0.210054    
   Mn18       1.0     0.489646     0.827088     0.789990    
   O1         1.0     0.916360     0.914765     0.876440    
   O2         1.0     0.936747     0.929374     0.295721    
   O3         1.0     0.916457     0.248646     0.876513    
   O4         1.0     0.935511     0.260482     0.295967    
   O5         1.0     0.914000     0.582479     0.877514    
   O6         1.0     0.938699     0.592930     0.296843    
   O7         1.0     0.250327     0.914834     0.876450    
   O8         1.0     0.270992     0.926379     0.297751    
   O9         1.0     0.246851     0.246531     0.881556    
   O10        1.0     0.273648     0.260743     0.296173    
   O11        1.0     0.247773     0.583169     0.880312    
   O12        1.0     0.275538     0.590629     0.295895    
   O13        1.0     0.584201     0.911488     0.879016    
   O14        1.0     0.603789     0.929464     0.295584    
   O15        1.0     0.584674     0.246418     0.881647    
   O16        1.0     0.605947     0.262056     0.288962    
   O17        1.0     0.584387     0.582435     0.877649    
   O18        1.0     0.603371     0.590808     0.295373    
   O19        1.0     0.086138     0.084060     0.122686    
   O20        1.0     0.062021     0.073483     0.703064    
   O21        1.0     0.083852     0.418276     0.123532    
   O22        1.0     0.064451     0.406349     0.703902    
   O23        1.0     0.083878     0.751377     0.123609    
   O24        1.0     0.063167     0.737350     0.704185    
   O25        1.0     0.415510     0.084244     0.122461    
   O26        1.0     0.396793     0.075706     0.704433    
   O27        1.0     0.414913     0.420399     0.118285    
   O28        1.0     0.393877     0.404763     0.710815    
   O29        1.0     0.415682     0.755226     0.121089    
   O30        1.0     0.396377     0.737253     0.704303    
   O31        1.0     0.752128     0.083434     0.119688    
   O32        1.0     0.724156     0.075781     0.704430    
   O33        1.0     0.753016     0.420350     0.118463    
   O34        1.0     0.726137     0.406342     0.704018    
   O35        1.0     0.749661     0.751850     0.123597    
   O36        1.0     0.728842     0.740398     0.702258    
   K1         1.0     0.500117     0.332853     0.500076    
\end{verbatim}
\normalsize

Transition state for \ce{Na+} intercalation
\footnotesize
\begin{verbatim}
_cell_length_a                         8.612171
_cell_length_b                         8.612713
_cell_length_c                         10.616097
_cell_angle_alpha                      69.567429
_cell_angle_beta                       74.300522
_cell_angle_gamma                      59.973324
_cell_volume                           634.387301
_space_group_name_H-M_alt              'P 1'
_space_group_IT_number                 1

loop_
_space_group_symop_operation_xyz
   'x, y, z'

loop_
   _atom_site_label
   _atom_site_occupancy
   _atom_site_fract_x
   _atom_site_fract_y
   _atom_site_fract_z
   Mn1        1.0     0.858838     0.132411     0.228249    
   Mn2        1.0     0.809916     0.201271     0.768424    
   Mn3        1.0     0.857720     0.465479     0.231155    
   Mn4        1.0     0.810265     0.535295     0.765474    
   Mn5        1.0     0.857035     0.798141     0.233222    
   Mn6        1.0     0.809737     0.867635     0.767620    
   Mn7        1.0     0.189657     0.131400     0.234681    
   Mn8        1.0     0.142556     0.201257     0.768554    
   Mn9        1.0     0.190438     0.465047     0.231655    
   Mn10       1.0     0.141005     0.534551     0.771604    
   Mn11       1.0     0.190286     0.798609     0.232538    
   Mn12       1.0     0.143161     0.868518     0.766690    
   Mn13       1.0     0.523602     0.132269     0.228470    
   Mn14       1.0     0.473814     0.195190     0.784831    
   Mn15       1.0     0.525784     0.471913     0.214954    
   Mn16       1.0     0.476064     0.534549     0.771574    
   Mn17       1.0     0.523185     0.799038     0.232261    
   Mn18       1.0     0.476936     0.867662     0.767736    
   O1         1.0     0.905859     0.945018     0.863835    
   O2         1.0     0.952360     0.877119     0.329465    
   O3         1.0     0.905699     0.278637     0.863775    
   O4         1.0     0.950232     0.209416     0.329325    
   O5         1.0     0.903678     0.612384     0.864727    
   O6         1.0     0.953599     0.540830     0.330375    
   O7         1.0     0.239112     0.945033     0.863903    
   O8         1.0     0.286229     0.873956     0.330893    
   O9         1.0     0.233827     0.277273     0.871362    
   O10        1.0     0.287729     0.209275     0.329861    
   O11        1.0     0.237389     0.613279     0.866678    
   O12        1.0     0.293093     0.538855     0.326911    
   O13        1.0     0.573675     0.939123     0.869465    
   O14        1.0     0.618094     0.877124     0.329441    
   O15        1.0     0.574060     0.277216     0.871199    
   O16        1.0     0.618929     0.214385     0.321917    
   O17        1.0     0.573502     0.612225     0.864826    
   O18        1.0     0.617065     0.539288     0.326303    
   O19        1.0     0.096333     0.054001     0.135519    
   O20        1.0     0.047156     0.125617     0.669402    
   O21        1.0     0.094669     0.387982     0.136190    
   O22        1.0     0.049677     0.457410     0.670516    
   O23        1.0     0.094512     0.720955     0.136287    
   O24        1.0     0.047446     0.789714     0.670459    
   O25        1.0     0.426429     0.054472     0.135322    
   O26        1.0     0.383110     0.127399     0.673464    
   O27        1.0     0.425436     0.389665     0.128830    
   O28        1.0     0.380812     0.452636     0.677757    
   O29        1.0     0.426324     0.727640     0.130531    
   O30        1.0     0.382012     0.789723     0.670540    
   O31        1.0     0.762271     0.053510     0.133378    
   O32        1.0     0.706626     0.127455     0.673378    
   O33        1.0     0.765996     0.389748     0.128596    
   O34        1.0     0.711981     0.457507     0.670461    
   O35        1.0     0.760790     0.721854     0.136197    
   O36        1.0     0.713687     0.792606     0.669106    
   Na1        1.0     0.500538     0.332769     0.500112    
\end{verbatim}
\normalsize

\end{document}